\documentclass[10pt,conference]{IEEEtran}
\IEEEoverridecommandlockouts

\usepackage{rotating}
\usepackage{enumitem}
\usepackage{booktabs}
\usepackage{multirow}
\usepackage{graphicx}
\usepackage{bigstrut}
\usepackage[noadjust]{cite}
\usepackage[linesnumbered,ruled,vlined, noend]{algorithm2e}
\usepackage{mathtools}
\usepackage{multirow}
\usepackage{booktabs} 
\usepackage{graphicx}
\usepackage[belowskip=-12pt]{caption}
\usepackage{subcaption}
\usepackage{rotating}
\usepackage{xspace}
\usepackage{framed}
\usepackage{syntax}
\usepackage{parcolumns}
\usepackage{pifont}
\usepackage{adjustbox}
\usepackage{array}
\usepackage{float}
\usepackage{tabularx}
\usepackage{xcolor}
\usepackage{stfloats}
\usepackage{wrapfig}
\usepackage{listings}
\usepackage{color, colortbl}
\usepackage{balance} 
\usepackage{lscape}
\usepackage{hyperref}
\usepackage{url}
\usepackage{cleveref}
\usepackage{soul}
\usepackage[tikz]{bclogo}
\usepackage{soul}
\usepackage{makecell}
\usepackage[many]{tcolorbox}
\usepackage{bigstrut}

\definecolor{problemblue}{RGB}{100,134,158}
\definecolor{idiomsgreen}{RGB}{0,162,0}
\definecolor{exercisebgblue}{rgb}{0,  .69,  .941}
\definecolor{deepgreen}{rgb}{0.0, 0.5, 0.0}
\definecolor{codegreen}{rgb}{0,0.6,0}
\definecolor{codegray}{rgb}{0.5,0.5,0.5}
\definecolor{codepurple}{rgb}{0.58,0,0.82}
\definecolor{backcolour}{rgb}{0.95,0.95,0.92}
\definecolor{redColor}{RGB}{255,0,0}
\definecolor{Gray}{gray}{0.1}
\definecolor{javared}{rgb}{0.6,0,0} 
\definecolor{javagreen}{rgb}{0.25,0.5,0.35} 
\definecolor{javapurple}{rgb}{0.5,0,0.35} 
\definecolor{javadocblue}{rgb}{0.25,0.35,0.75} 
\definecolor{ibmblue}{RGB}{63,97,246}
\definecolor{maroon}{RGB}{105,33,61}
\definecolor{teal}{RGB}{59,115,115}
\crefformat{section}{\S#2#1#3} 
\crefformat{subsection}{\S#2#1#3}
\crefformat{subsubsection}{\S#2#1#3}
\crefformat{figure}{Fig.~#2#1#3}
\crefformat{table}{Tab.~#2#1#3}
\crefformat{algorithm}{Alg.~#2#1#3}

\newcommand{\aster}{\textsc{aster}\xspace}

\NewDocumentCommand{\rangeet}
{ mO{} }{\textcolor{blue}{\textsuperscript{\textit{rangeet}}\textsf{\textbf{\small[#1]}}}}
\NewDocumentCommand{\raju}
{ mO{} }{\textcolor{red}{\textsuperscript{\textit{raju}}\textsf{\textbf{\small[#1]}}}}
\NewDocumentCommand{\soo}
{ mO{} }{\textcolor{purple}{\textsuperscript{\textit{Soo}}\textsf{\textbf{\small[#1]}}}}
\NewDocumentCommand{\rahul}
{ mO{} }{\textcolor{magenta}{\textsuperscript{\textit{rahul}}\textsf{{\small[#1]}}}}

\newcommand{\etal}{{\em et al.}\xspace}
\newcommand*\circled[2][fill=black]{\tikz[baseline=(char.base)]{
    \footnotesize
    \node[shape=circle, #1, inner sep=1pt] (char) {\textcolor{white}{#2}};}}
\newenvironment{findingBox}[2]{%
	\begin{tcolorbox}[colframe=black!90,colback=exercisebgblue!10,boxrule=.5pt, left=0pt, right = 0pt, top=0pt, bottom=0pt]{\textbf{Finding #1:} #2} 
}{%
	\end{tcolorbox}
}

\newcommand\bi{\begin{itemize}[leftmargin=*, itemindent=0pt, wide=0pt]}
\newcommand\ei{\end{itemize}}

\makeatletter
\newcommand{\removelatexerror}{\let\@latex@error\@gobble}
\makeatother

\AtBeginDocument{%
	\providecommand\BibTeX{{%
			\normalfont B\kern-0.5em{\scshape i\kern-0.25em b}\kern-0.8em\TeX}}}

\title{ASTER: Natural and Multi-language Unit Test Generation with LLMs}

\begin{document}
\author{\IEEEauthorblockN{Rangeet Pan\IEEEauthorrefmark{1}\textsuperscript{$1$},
Myeongsoo Kim\IEEEauthorrefmark{2}\IEEEauthorrefmark{3}\textsuperscript{$2$},
Rahul Krishna\IEEEauthorrefmark{1}\textsuperscript{$3$},
Raju Pavuluri\IEEEauthorrefmark{1}\textsuperscript{$4$}, 
Saurabh Sinha\IEEEauthorrefmark{1}\textsuperscript{$5$}}
\IEEEauthorblockA{\IEEEauthorrefmark{1}IBM Research, 
Yorktown Heights, NY, 10598, USA.\\
\IEEEauthorrefmark{2}Georgia Tech, Atlanta, GA, 30332, USA.\\
\{{\textsuperscript{$1$}rangeet.pan, \textsuperscript{$3$}rkrsn\}@ibm.com, \{\textsuperscript{$4$}pavuluri, \textsuperscript{$5$}sinhas\}@us.ibm.com, \textsuperscript{$2$}mkim754@gatech.edu}}
\thanks{
\scriptsize\IEEEauthorrefmark{3}~Myeongsoo Kim was an intern at IBM Research.}
}

\maketitle
\thispagestyle{plain}
\pagestyle{plain}
\begin{abstract}
	Implementing automated unit tests is an important but time-consuming activity in software development.
To assist developers in this task, many techniques for automating unit test generation have been developed. However, despite this effort, usable tools exist for very few programming languages.
Moreover, studies have found that automatically generated tests suffer poor readability and do not resemble developer-written tests. In this work, we present a rigorous investigation of how large language models (LLMs) can help bridge the gap. We describe a generic pipeline that incorporates static analysis to guide LLMs in generating compilable and high-coverage test cases. We illustrate how the pipeline can be applied to different programming languages, specifically Java and Python, and to complex software requiring environment mocking. We conducted an empirical study to assess the quality of the generated tests in terms of code coverage and test naturalness---evaluating them on standard as well as enterprise Java applications and a large Python benchmark. Our results demonstrate that LLM-based test generation, when guided by static analysis, can be competitive with, and even outperform, state-of-the-art test-generation techniques in coverage achieved while also producing considerably more natural test cases that developers find easy to understand.
We also present the results of a user study, conducted with 161 professional developers, that highlights the naturalness characteristics of the tests generated by our approach. 
\end{abstract}
\section{Introduction}
\label{sec:intro}

Unit testing is a key activity in software development that serves as the first line of defense against the introduction of software bugs. Manually writing high-coverage unit tests can be tedious and time consuming.  To address this, many automated test generation (ATG) techniques have been developed, aimed at reducing the cost of manual test suite development: over the last few decades, this research has produced a variety of approaches based on symbolic analysis (e.g.,~\cite{visser:2004, cadar:2008, pasareanu:2010, godefroid2005dart, sen2005cute, xie:2005, tillmann2008pex}), search-based techniques (e.g.,~\cite{tonella:2008, harman:2010:tse, mcminn:2004, fraser2011evosuite, lin:2021, lukasczyk2022pynguin}), random and adaptive-random techniques (e.g.,~\cite{pacheco2007feedback, ciupa:2008:icse, chen:2010:jss, lin:2009:ase, arcuri:2011:issta, lukasczyk:2023:emse}), etc.

\begin{figure}[!htp]
    \centering
    \includegraphics[width=\linewidth,trim = {1.4cm 0 2.5cm 1cm}]{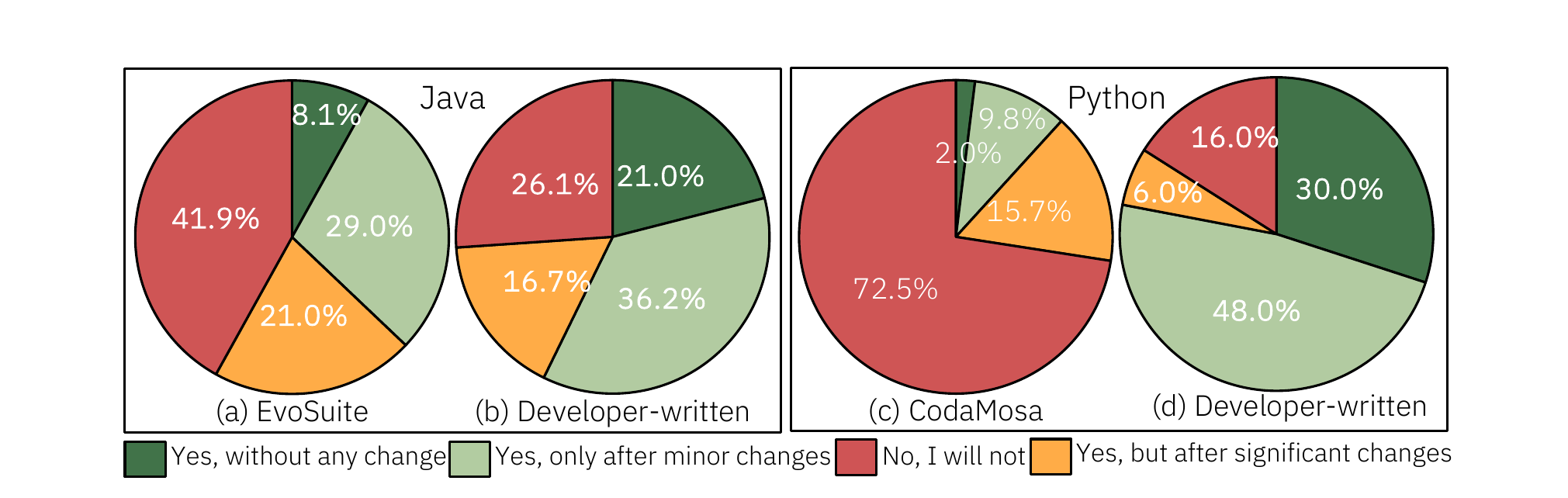}
    \vspace{-18pt}
    \caption{\small Results of the survey question on whether developers would add automatically generated tests to regression test suites. 
    }
    \vspace{6pt}
    \label{fig:surveyq18-intro}
\end{figure}

These techniques have achieved considerable success in generating high-coverage test suites with good fault-detection ability, but they still have several key limitations---with respect to test readability, test scenarios covered, and test assertions created.  Previous studies (e.g.,~\cite{fraser:2015}) have shown that developers find automatically generated tests lacking in these characteristics, suffering poor readability and comprehensibility, covering uninteresting sequences, and containing trivial or ineffective assertions. Automatically generated tests are also known to contain anti-patterns or test smells~\cite{panichella:2020:testsmells} and generally not perceived as being ``natural'' in the sense that they do not resemble the kinds of tests that developers write. All these factors inhibit the adoption of ATG tools in practice, as developers consider the tests generated by these tools to be hard to maintain
and are reluctant to add them to regression test suites without some or considerable rewrite.

To understand developer perception of tests created by ATG tools, we conducted a survey of professional software developers (details in \S\ref{subsec:rq2naturalness}). \cref{fig:surveyq18-intro} shows the results for one of the survey questions---on readiness of automatically generated tests (by two ATG tools, EvoSuite~\cite{evosuite} for Java and CodaMosa~\cite{codamosa} for Python) for addition to regression test suites, comparing them with human-written tests. For EvoSuite, 42\% of the responses to this question stated that the tests were not suitable for addition to a regression test suite, with an additional 21\% stating that the tests could be added only after significant modifications. For CodaMosa, the result is worse, with 87\% of the responses falling in these two categories. These values are significantly lower for human-written tests, which shows that ATG tools need considerable improvement in this respect---so that they generate tests that are readily usable by developers in practice. 

In this work, we investigate the potential of LLMs in generating natural test cases that developers consider to be readily usable for building regression test suites. Our goal is to leverage the inherent ability of LLMs in synthesizing natural-looking code, while also avoiding the pitfalls of using LLMs off-the-shelf for test generation, which can result in the creation of tests that often have compilation errors or achieve low code coverage. Moreover, by leveraging LLMs' intrinsic knowledge of different programming languages (PL) and frameworks, our broader goal is to build a multi-language ATG tool that performs well on enterprise-grade applications, with their complex structure, framework dependencies, and multi-tiered architecture. Finally, we aim to develop a technique in which the LLM is a pluggable component and, therefore, can be configured to work with different language models.

\begin{figure*}
    \centering
    \includegraphics[width=.83\linewidth]{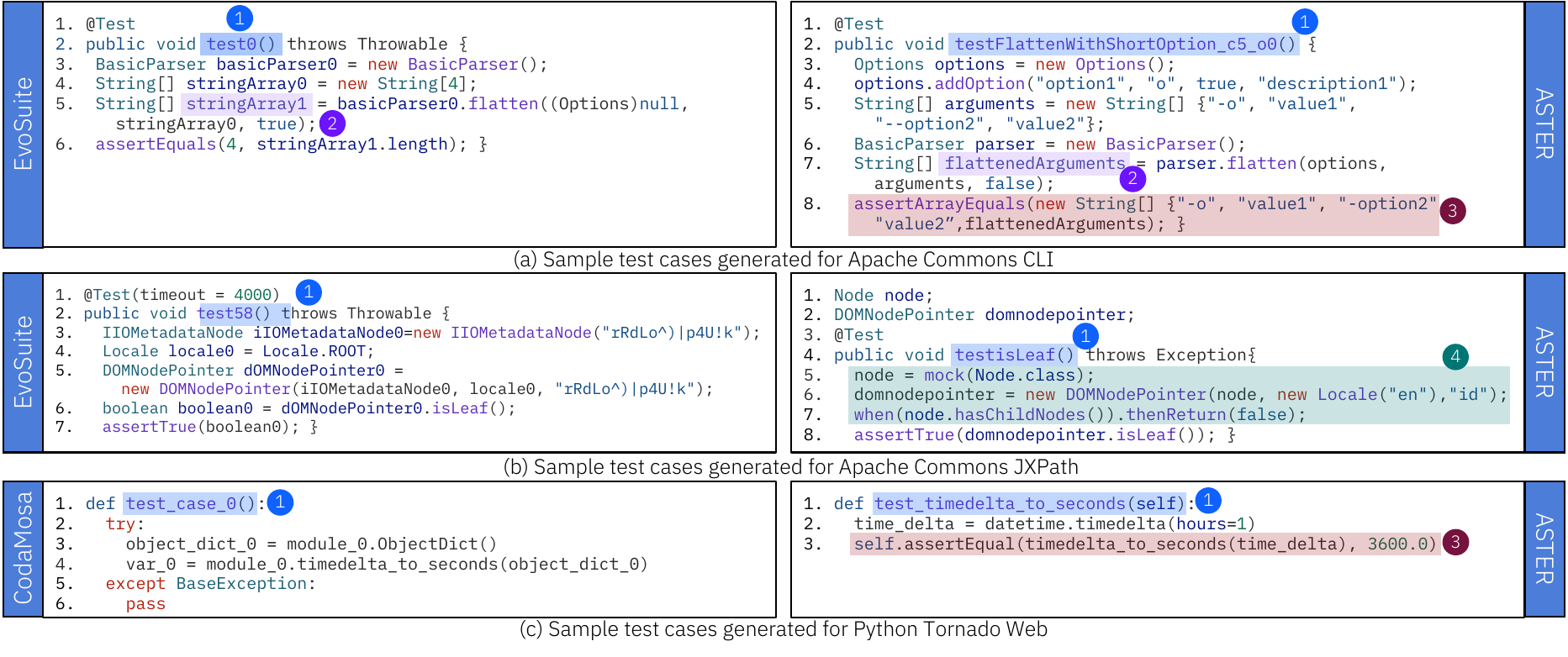}
    \vspace{-8pt}
    \caption{\small Illustration of naturalness (in terms of test names, variable names, and assertions) and mocking in test cases generated by the LLM-assisted technique of \aster (right) compared with tests generated by EvoSuite~\cite{fraser2011evosuite} and CodaMosa~\cite{lemieux2023codamosa} (left).}
    \vspace{-5pt}
    \label{fig:naturlaness-examples}
\end{figure*}

We present a technique for LLM-assisted test generation guided by program analysis. The technique consists of preprocessing and postprocessing phases that wrap LLM interactions.  The \textit{preprocessing phase} performs static program analysis to compute relevant information to be included in LLM prompts for a given method under test (or \textit{focal method}). This ensures that the LLM prompt has sufficient context (similar to the information that a developer would use while writing test cases for a method) and increases the chances that it generates compilable and meaningful test cases. The \textit{postprocessing phase} checks the generated tests for compilation and runtime errors and constructs new prompts aimed at fixing the errors. After the test-repair step, the technique produces a set of passing tests and a set of failing tests for the focal method. To increase code coverage, the technique includes a \textit{coverage-augmentation phase}, in which prompts are crafted for instructing the LLM to generate test cases aimed at exercising uncovered lines of code.

We implemented the technique in a tool, called \aster, for two programming languages, Java and Python, thus demonstrating the feasibility of building multilingual unit test generators with LLMs guided by lightweight program analysis.
\aster also incorporates mocking capability for Java unit test generation, which makes it applicable to applications that perform database operations, implement services, or use complex libraries. We present a generic approach for generation of test with mocks that is extensible to different library APIs.

\textit{Results-at-a-glance.}
We performed a comprehensive evaluation of \aster, using six general-purpose and code LLMs, to assess the generated tests in terms of code coverage and test naturalness; we also compared \aster against state-of-the-art Java and Python unit test generators, EvoSuite~\cite{evosuite} and CodaMosa~\cite{codamosa}. Our results show that, in terms of code coverage, \aster is very competitive with EvoSuite (+2.0\%, -0.5\%, and +5.1\% line, branch, and method coverage) for Java SE applications while performing significantly better (+26.4\%, +10.6\%, and +18.5\%, line, branch, and method coverage) for Java EE applications. For Python, \aster outperforms CodaMosa considerably (+9.8\%, +26.5\%, and +22.5\% line, branch, and method coverage). With respect to test naturalness, our quantitative evaluation shows that \aster-generated tests have superior naturalness characteristics than EvoSuite- and CodaMosa-generated tests, while our user survey (of 161 professional software engineers) confirms strong developer preference for \aster test cases over the tests generated by the other tools: e.g., 70\% and 88\% of the responses for Java and Python, respectively, stated that \aster-generated tests could be added to a regression suite with no or only minor changes.

Our work corroborates and strengthens the findings on high developer acceptance rates of LLM-generated tests reported in recent industry studies~~\cite{meta2024} by providing additional evidence in a different industry setting. Our work also provides new insights via a more rigorous evaluation of LLM-based test generation, showing that smaller models, when provided with suitable code context, can be made competitive to larger models and even outperform those models.

\aster's Java test generation capability is offered in the IBM watsonx Code Assistant for Enterprise Java Applications product~\cite{ibm:wca4eja}.
The results of our experiments and the naturalness evaluator are available in our artifact~\cite{aster:artifact}.


The main contributions of this work include:
\begin{itemize}[leftmargin=*]
    \item An LLM-assisted test-generation technique, and its implementation in a tool called \aster, that is fueled by static analysis and can be extended to multiple PLs.
    \item Empirical results showing that \aster for Java and Python (configured with six different LLMs) is competitive with, and even outperforms, state-of-the-art ATG tools in code coverage achieved while generating considerably more natural tests than those tools.  
    \item In-depth quantitative and qualitative analysis of test naturalness via an automated approach and a survey of 161 professional developers highlighting the naturalness attributes of \aster-generated tests.
\end{itemize}

\section{Motivation}
\label{sec:motivation}

The primary motivation for LLM-assisted test generation is to overcome the limitation of conventional ATG tools w.r.t lack of naturalness in the tests they create. By leveraging the LLM's inherent ability of creating natural-looking code, we can generate more readable, comprehensible, and meaningful test cases. 
Another motivation is to build multi-lingual unit test generators---leveraging the LLM's understanding of the syntax and semantics of multiple PLs on which the models are typically trained. Building such test generators using conventional approaches (e.g., symbolic or evolutionary techniques) can be challenging and no such test generators exist. 
In contrast, with lightweight static analysis guiding LLM interactions, a multi-lingual unit test generator can be easily implemented. 
Finally, an LLM-based approach can also address test generation for complex applications, which often require mocking.

To illustrate these benefits, \cref{fig:naturlaness-examples} presents sample \aster-generated test cases and tests generated by two existing ATG tools: EvoSuite~\cite{fraser2011evosuite} for Java and CodaMosa~\cite{lemieux2023codamosa} for Python. 
The \aster-generated test cases have more meaningful test names\,\circled[fill=ibmblue]{1}, variable names\,\circled[fill=codepurple]{2}, and assertions\,\circled[fill=maroon]{3} than the EvoSuite- or CodaMosa-generated tests. For instance, consider the test cases for Apache Commons CLI~\cite{commonscli}. The variable storing the return value from \texttt{\small flatten()} is called \texttt{\small flattenedArguments} in the \aster test---clearly capturing the meaning of the stored data---whereas the corresponding variable in the EvoSuite test is called \texttt{\small stringArray1} (line~5), which captures simply the data structure type instead of any meaning of the stored data.  
Similarly, the Python test case generated by \aster, shown in \cref{fig:naturlaness-examples}(c), has meaningful test name and variable names. Moreover, the assertion in the test case is generated taking into account the expected transformation of the input by the focal method, converting the input hour value to minutes.
\cref{fig:naturlaness-examples}(b) shows a unit test case with mocking of library APIs generated for Apache Commons JXPath~\cite{commonsjxpath}. The test mocks the behavior of the \texttt{\small org.w3c.dom.Node} library class\,\circled[fill=teal]{4}.

\vspace{-3pt}
\section{Our Approach}
\label{sec:methodology}

\begin{figure}
    \centering
\includegraphics[width=\columnwidth]{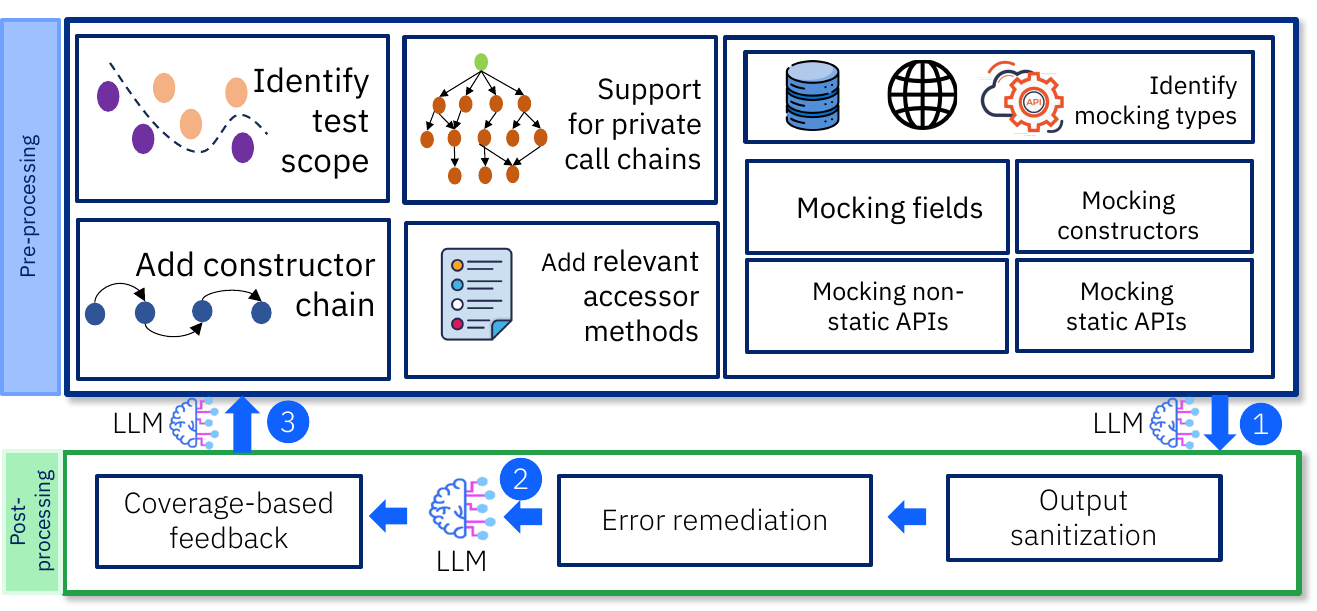}
    \vspace{-16pt}
    \caption{\small Overview of \aster. \textcircled{\raisebox{-.9pt} {1}}, \textcircled{\raisebox{-.9pt} {2}}, \textcircled{\raisebox{-.9pt} {3}} represent test-generation, test-repair, and coverage-augmentation prompts.}
    \label{fig:overview}
    \vspace{7pt}
\end{figure}

\begin{figure*}[htbp!]
    \centering
    \includegraphics[width=0.91\linewidth]{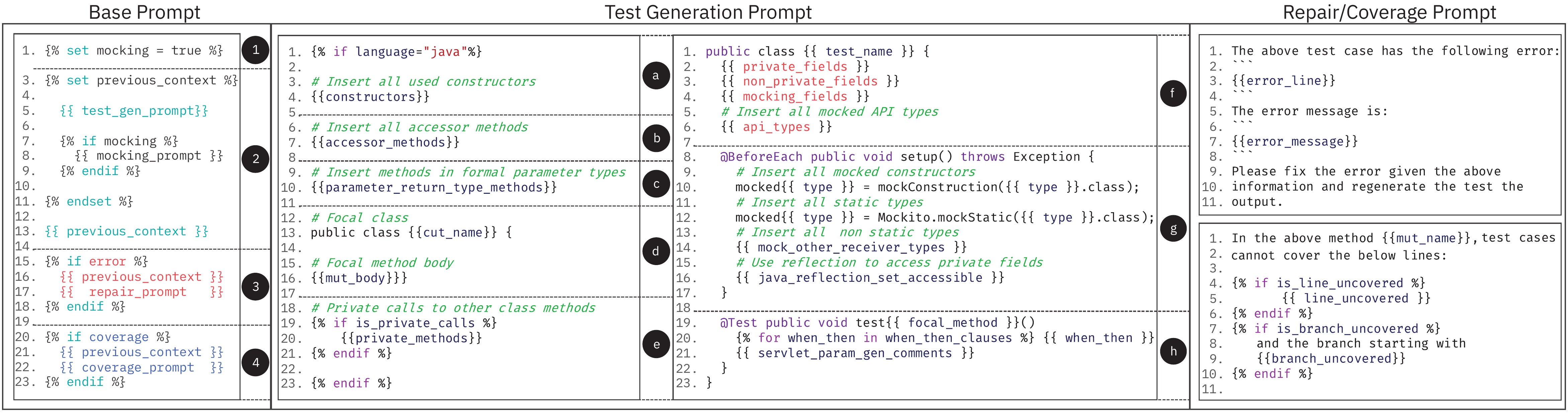}
    \vspace{-5pt}
    \caption{\small Templates for composing prompts for test generation, test repair, and coverage augmentation.}
    \vspace{-7pt}
    \label{fig:prompting-templates}
\end{figure*}

Figure~\ref{fig:overview} illustrates our LLM-based test-generation technique, guided by program analysis. The process consists of two phases, both utilizing stage-specific prompting: (1)~\textit{Preprocessing}, in which static analysis extracts context and constructs initial prompts (\cref{fig:prompting-templates}) for LLM-driven unit test generation in Java and Python and (2)~\textit{Postprocessing}, where the generated tests are repaired for compilation and runtime errors, and augmented for coverage enhancement. 

\paragraph{Preprocessing}
\label{subsec:preprocessing}

The preprocessing phase performs static analysis of the application to collect relevant context that an LLM might require to generate unit tests. The key objective of this stage is to gather all the necessary information pertaining to the focal method and its broader context within the application. This information is used to populate fields \circled{1} and \circled{2} of the prompt template shown in Figure~\ref{fig:prompting-templates}. In this section, we discuss the preprocessing steps for Java and how those steps help construct a prompt for the LLM. 

\vspace{-5pt}
\bi
\item[~]\textit{Testing scope.}~%
The first step in the preprocessing phase identifies the testing scope given a Java focal class $f_c$. The testing scope lists the set of focal methods to be targeted for test generation. This set consists of (1) public, protected, and package-visibility methods declared in $f_c$ and (2) any inherited method implementations from an abstract super class of $f_c$. The set excludes inherited methods from a non-abstract super class as those methods are targeted for test generation in the context of their declaring class as the focal class. If $f_c$ is an abstract class, the testing scope consists of static methods, if any, declared in $f_c$. 
The focal class and method are used to populate part \circled{d} of the test-generation prompt.

\item[~]\textit{Relevant constructors.}~\aster next identifies the relevant constructors for a focal method so that their signatures can be specified to the LLM, enabling it to create required objects for invoking the focal method. The relevant constructors for a focal method $m$ include the constructors of the focal class (if $m$ is a virtual method), along with the constructors of the each formal parameter type of $m$, considering application types only (i.e., ignoring library types). The analysis is done transitively for the formal parameter types of each identified constructor, thus ensuring that the LLM prompt includes comprehensive context information about how to create instances of application types that may be needed for testing $m$. 
The discovered constructors are used in the test-generation prompt (\cref{fig:prompting-templates}~\circled{a}).

\item[~]\textit{Relevant auxiliary methods.}~%
Accessor methods, which consist of getters and setters, provide a mechanism for reading and modifying internal object state, while preserving encapsulation. Our approach identifies setter methods in the focal class and in each formal parameter type of the focal method (if the type is an application class). Signatures of these setters are added to the LLM prompt to help with setting up suitable object states for invoking the focal method. For reading object state, our approach computes getters of the focal class and the return type of the focal method (limiting to application return types), and includes their signatures in the LLM prompt. This can help the LLM generate suitable assertion statements for verifying relevant object state (of the receiver object or the returned object) after the focal method call.
This information is used to fill section \circled{b} of the test-generation prompt in \cref{fig:prompting-templates}.

\item[~]\textit{Private methods.}~
Private methods are inaccessible outside the class. One way to test them is by using reflection (in the case of Java) to overwrite method accessibility. However, that is considered a bad practice. A more accepted approach is to test private methods indirectly by invoking them through non-private methods of the class.  To facilitate this, we compute the class call graph of the focal class, identify call chains from non-private methods to private methods, and provide these call chains to the LLM to enable it to generate test cases that invoke private methods through externally visible methods. 
This information is used to populate ~\cref{fig:prompting-templates} \circled{e} section.

\item[~]\textit{Facilitating Mocking.}~
Mocking is a technique in software testing that allows developers to create simulated objects mimicking the behavior of actual components. This is particularly useful for simulating external services, databases, or third-party API calls. By incorporating mocking capabilities, LLM-generated tests can verify code behavior and ensure proper component interaction without needing to understand the full implementation details of all dependencies in the application. Our approach uses a systematic method to identify relevant candidates for mocking.
The mocking facilitation process consists of two main steps: identifying fields and types to be mocked, and identifying methods to be stubbed. In the first step, we examine the focal class and method to discover all candidate fields and types that need to be mocked. We begin by iterating through the fields defined in the focal class, identifying each field whose type matches one of the mockable APIs. We then perform a comprehensive search starting from the formal parameter types of the focal method and class. This search is transitive, meaning we also consider the parameter types of constructors for each identified type.
The identified fields and types to be mocked are used to populate section \circled{f} of the test-generation prompt (\cref{fig:prompting-templates}).
The second step involves determining the scope for creating method stubs. These stubs emulate method calls using ``when'' and ``then'' clauses. ``When'' clauses define the conditions under which a mock should return a specified value, while ``then'' clauses specify the expected behavior once these conditions are met. To identify the stubbing scope, we start with the focal method and its class constructors. We then include all methods reachable from the focal method within the focal class. If the focal class is a service entry class, we also include its overridden methods. Finally, we examine all call sites within this scope, categorizing them as mockable constructor calls, static calls, or API calls based on their types. These information are used to populate parts \circled{g} and \circled{h} of the test-generation prompt (\cref{fig:prompting-templates}).

\item[~]\textit{Preprocessing for Python.}~
Unlike Java, Python's more flexible structure, where a single file or module can contain a mix of functions, classes, and standalone code, lends itself better to module-scoped test generation. Therefore, for Python test generation, \aster targets a module as a whole for test generation, including all classes, methods, and functions declared in the module.
This module-level approach together with distinct language features of Python allows us to omit some preprocessing steps that are necessary for Java but irrelevant for Python. Because our approach provides the entire module to the LLM, instructing the LLMs about calls to private methods is unnecessary. Member visibility is Python is specified via naming conventions (names beginning with single underscore for protected members and double underscores for private members) and there is no strict encapsulation as these members can be accessed via name mangling. Python's properties feature (using \texttt{\small @property} decorators) also lets private members to be accessed directly. Thus, identification of accessors is also unnecessary for Python test generation. In terms of relevant constructors, the constructor definitions in the focal module are already available to the LLM. Additionally, we add constructors for all imported modules to the LLM prompt. Also, we found that, for Python, adding a few examples of tests helps LLMs produce more predictable output.
For Java, this was not necessary, but investigation of RAG-based approaches for incorporating in-context learning for test generation is an interesting future research direction.

\ei







\paragraph{Postprocessing}
\label{subsec:postprocessing}

Despite using rich context gathered through static analysis, LLMs can still generate code that: (1) does not adhere to PL syntax (non-parsable), (2) does not adhere to the testing framework, (3) does not have sufficient details such as imports, package name, class name, folder location (for Java tests, it is essential to follow the source directory structure), (4) has compilation issues, or (5) runs but fails due to assertion or runtime failures. During postprocessing, we perform several fixing steps to remediate these issues. Additionally, we perform coverage augmentation to increase the coverage of the initial set of tests generated by \aster.

\subsubsection{Output sanitization}
First, we sanitize the generated code for extraneous content (e.g., natural language text) and ensure the syntactic correctness of the generated test cases. We also gather the required imports by static analysis. For instance, we look into the focal class and its superclasses to gather the used imports. We add all application classes as imports because, often, test cases refer to other application classes that may not be imported in the focal class or its superclasses. We also add a few imports related to the testing framework. Any unused imports added in this process is removed at the end of test generation. Next, for Java, we augment the build file to add the testing framework dependencies, if required, and identify the Java and JUnit versions from the build file, which are used for generating test cases. Third, we check the names of the test and setup/teardown methods to avoid duplicate content as the LLM may generate test cases without paying attention to details. Fourth, we add the test framework details, such as annotations, timeout values, etc., and also add exception information. Finally, we form the test class for each focal class and store it to desired location with all essential details. These steps are very crucial as, even with the best LLM, without code sanitization, a significant proportion of the generated tests may end up with compilation or runtime errors.

\subsubsection{Error remediation}
After sanitizing the code, we perform compilation and runtime checks. For each type of error, we first apply a set of rule-based fixes. For instance, we try to resolve errors due to naming clash of methods, variables, etc.; we attempt to fix assertion failures where the expected and actual values do not match. 
Also, we localize the error, decompose tests with multiple assertions into multiple tests with single assertions, and fix them. At the end of test generation, we merge the tests back to reduce the number of test cases. Then, for the rest of the errors, we gather all the context required for fixing the error, such as error message from compiler feedback, erroneous line, and context surrounding the focal method and re-prompt LLM for fixing the test cases (depicted in \cref{fig:prompting-templates}-\circled{3}). We sanitize the fixed code and add them to the test suite.

\subsubsection{Increasing coverage of generated tests}
Error-free test cases are executed to measure code coverage, which is further analyzed to identify uncovered code lines and update the LLM prompt template (\cref{fig:prompting-templates}) to guide the LLM to generate test cases targeting the uncovered code lines. We also add the context gathered in the preprocessing phase to the prompt. 

\textit{Postprocessing for Python.} 
For Python, we use Pylint~\cite{pylint}, for identifying compilation and parsing errors. 
Specifically, \aster focuses on the error category of Pylint warnings and uses this information to guide LLMs in fixing compilation and parsing issues. Then, tests are executed using Pytest~\cite{pytest} and the output is used to provide feedback to the LLMs in case of failures. 
Finally, Coverage.py~\cite{coverage} is used to measure coverage and perform the coverage augmentation.

\begin{table}
\centering
  \setlength{\tabcolsep}{2pt}
    \caption{\small Models used in the evaluation.}
    \vspace{-3pt}
\resizebox{0.92\linewidth}{!}{
\begin{tabular}{lccccc}
\toprule
\textbf{Model Name} & \textbf{Provider} & \textbf{Update Date} & \textbf{Model Size} & \textbf{License} & \textbf{Data Type}                                                                                                                                                                                                                                             \bigstrut
\\ 
\midrule
GPT-4-turbo               & OpenAI                                & May-24                                & 1.76T$^\dagger$     & Closed-source   & Generic                  
\bigstrut\\ 

Llama3-70b          & Meta                                   & Apr-24                                & 70B   & Llama-3 License*      & Generic                                       
\bigstrut
\\

\hline

CodeLlama-34b      & Meta                                   & Aug-23                                & 34B  & Llama-2 License*     & Code                                
\bigstrut\\
Granite-34b      & IBM                                   & May-24                                & 34B      & Apace 2.0 License*    & Code                             
\bigstrut\\
\hline

Llama3-8b      & Meta                                   & Apr-24                                & 8B    & Llama-3 License*    & Generic                               
\bigstrut\\
Granite-8b      & IBM                                   & May-24                                & 8B     & Apache 2.0 License*     & Code                             
\bigstrut\\ 
\bottomrule
\end{tabular}}\\
\tiny *Open-source with different licenses. $^\dagger$Model size not confirmed.
\label{tab:models}
\end{table}

\section{Experiment Setup}
\label{sec:setup}

\vskip -2pt
\noindent\textit{A. Research Questions:}
Our evaluation focuses on three research questions.

\begin{description}
\item[RQ1:] How effective is \aster in code coverage achieved with different models?
\item[RQ2:] How do developers perceive \aster-generated tests in terms of their comprehensibility and usability?
\item[RQ3:] How natural are \aster-generated tests?
\end{description}

\begin{table}
  \centering
  \vspace{10pt}
  \setlength{\tabcolsep}{3pt}
  \caption{\small Java and Python datasets used in the evaluation.}
  \resizebox{.7\linewidth}{!}{
    \begin{tabular}{clccc}
    \toprule
&    \textbf{Dataset} & \textbf{Classes/Modules} & \textbf{Methods} & \textbf{NCLOC}         
\\ 
\midrule
\multirow{4}{*}{\rotatebox{90}{Java SE}} &    Commons CLI &    31   &   305    &   2498    \\ 
&    Commons Codec &    97   &    776   &      9681 \\
&    Commons Compress &    500   &  3650     &    43545   \\
&    Commons JXPath &   180    &   1502    &     20142  \\
    \midrule
\multirow{4}{*}{\rotatebox{90}{Java EE}} & CargoTracker &   107    &    482   &   5445   \\
&    DayTrader &   148    &  1067     &    11409   \\
&    PetClinic &    23   &    84   &    805   \\
&    App X &  140     &  2111     &   21655   \\
    \midrule
&    Python Dataset &  283     &   2216    &   38633    \\
    \bottomrule
    \end{tabular}%
    }  \label{tb:dataset}%
\end{table}%

\vskip 3pt
\noindent\textit{B. Baseline test-generation tools:}
We evaluated \aster against two state-of-the-art unit test generators. For Java, we used EvoSuite~\cite{evosuite}, specifically Release 1.2.0. For Python, we selected CodaMosa and used the latest version from its repository~\cite{codamosa}. CodaMosa is built on Pynguin~\cite{lukasczyk2022pynguin}, a search-based test-generation tool, and improves upon it by leveraging LLM-generated test cases to expand the search space on reaching coverage plateaus. The available version of CodaMosa works with the deprecated code-davinci-002 model; we, therefore, updated the tool to work with GPT-3.5-turbo-instruct, the recommended replacement model for code-davinci-002~\cite{openai:deprecations}.
\vskip 3pt
\noindent\textit{C. Models:}
We selected six models for evaluation, including LLMs of different sizes (ranging from 8B to over 1T parameters), open-source and closed-source LLMs, and LLMs from different model families (GPT~\cite{openaimodels}, Llama-3~\cite{llama3models}, and Granite~\cite{granitecodemodels}), considering general-purpose and code models. Table~\ref{tab:models} provide details of the selected models. 

\vskip 3pt
\noindent\textit{D. Datasets:}
Table~\ref{tb:dataset} lists the datasets used in the evaluation. The Java dataset is split into Java Standard Edition (SE) and Enterprise Edition (EE) applications. For Java SE, we selected four Apache Commons projects: Commons CLI~\cite{commonscli}, Commons Codec~\cite{commonscodec}, Commons Compress~\cite{commonscompress}, and Commons JXPath~\cite{commonsjxpath}. For Java EE, we used three open-source applications (CargoTracker~\cite{cargotracker}, DayTrader~\cite{daytrader}, PetClinic~\cite{petclinic}), covering different Java frameworks and a proprietary enterprise application (called ``App X'' for confidentiality). 

For Python, we started with 486 modules from the CodaMosa artifact~\cite{codamosa:artifact}. We ran CodaMosa on this dataset, but encountered crashes on 203 of the modules. We excluded those modules, resulting 283 modules from 20 projects. 
We went through these failed modules individually to check whether they were indeed failures.




\vskip 3pt
\noindent\textit{E. Evaluation metrics:} 

\textbf{Code coverage:}
We used JaCoCo~\cite{jacoco} for Java and Coverage.py~\cite{coverage} for Python to measure coverage. Because Coverage.py does not report method coverage, we implemented custom code for inferring method coverage from line coverage. 


\textbf{Naturalness:}
For test cases, we consider naturalness to encompass different characteristics, such as (1) readability in terms of meaningfulness of test and variable names, (2) quality of test assertions, (3) meaningfulness of test sequences, (4) quality of input values, and (5) occurrences of test smells or anti-patterns. Our evaluation focuses on assessing characteristics~1 and~2 quantitatively; additionally, we conducted a large-scale developer survey, which provides developer perspectives on characteristics 1--4. For studying occurrences of test smells, we attempted to use a test-smell detection tool for Java, TSDetect~\cite{tsdetect, peruma:2020:tsdetect}, but ran into numerous issues with the tool, such as identification of spurious test smells (false positives) and incorrect counts of test smells. We therefore chose to not use the tool, and perform our quantitative evaluation with a custom naturalness checker for characteristics (1) and (2) implemented using Tree-sitter~\cite{treesitter} and WALA~\cite{wala}. 
We note that our metrics for assessing test assertion quality include some of the test smells detected by TSDetect.

\paragraph*{Measuring assertion quality}
For assessing test assertion quality, our implementation uses the following metrics.
\begin{itemize}[wide = 0pt]
    \item \textit{Assertion ratio.} This measures the percentage of lines of code with assertions in a test case. Generally, tests with too many assertions can be brittle and break easily. 
    \item \textit{Tests with no assertions.} The percentage of test cases in a test file that have no assertions. Tests without assertions are considered an anti-pattern as they do not check for expected behavior and can only detect crashing failures. 
    \item \textit{Tests with duplicate assertions.} For a test file, the percentage of tests with assertions that contain duplicate assertions.
    \item \textit{Tests with null assertions.} For a test file, the percentage of tests with assertions that contain a null assertion, which is considered an anti-pattern. 
    \item \textit{Tests with exception assertions.} For a test file, the percentage of tests with assertions that contain an exception assertion. 
\end{itemize}



\paragraph*{Measuring meaningfulness of test method name}
A unit test name should clearly capture the functionality being tested, by including the focal method name and the functionality being tested if necessary. For example, a test that checks whether command-line options are correctly populated when long option values are provided can be named \texttt{\small test_getOptions_longOptions}. However, tools such as EvoSuite and CodaMosa often generate non-descriptive test names (e.g., \texttt{\small test01}) that do not reflect the functionality being tested.
We developed the following approach for measuring the meaningfulness of test names. 
First, for a given test file, we identify the focal class, similar to the Method2Test approach~\cite{tufano2020unit}.
For Python, we skip this step because focal methods can exist inside or outside of classes, and resolving imports is more complex. Instead, we directly identify focal methods, assuming any function call could be a focal method.
For Java, we examine all call sites in the test case, matching them with testable methods in the focal classes. This step results in a list of potential focal methods. We then check if one of these methods is mentioned in the test name. If it is, we assign a 50\% score for the match.
Next, we tokenize the remaining part of the test name after removing ``test'' keywords. We use camel-case and underscore splitting and generate all possible word combinations by merging them sequentially. For example, \texttt{\small test_addOption_longArgs_throwsException} is first matched with the focal method \texttt{\small addOption}, then the name is broken into \texttt{\small long}, \texttt{\small args}, and \texttt{\small longArgs}. For exceptions, we match them separately with thrown exceptions. We calculate the closeness score using the Levenshtein distance by matching tokens with code identifiers.

\paragraph*{Measuring meaningfulness of variable names in tests}
Consider the following example, showing two test cases that have the same set of steps.
Our premise is that the name of a variable of a data structure type (e.g., \texttt{\small String}, \texttt{\small List}) can be meaningful based on its context, whereas the meaningfulness of a variable name for a non-data structure type depends on both type and context.
\vspace{-5pt}
\begin{figure}[H]
		\centering
		\includegraphics[, ,width=\linewidth]{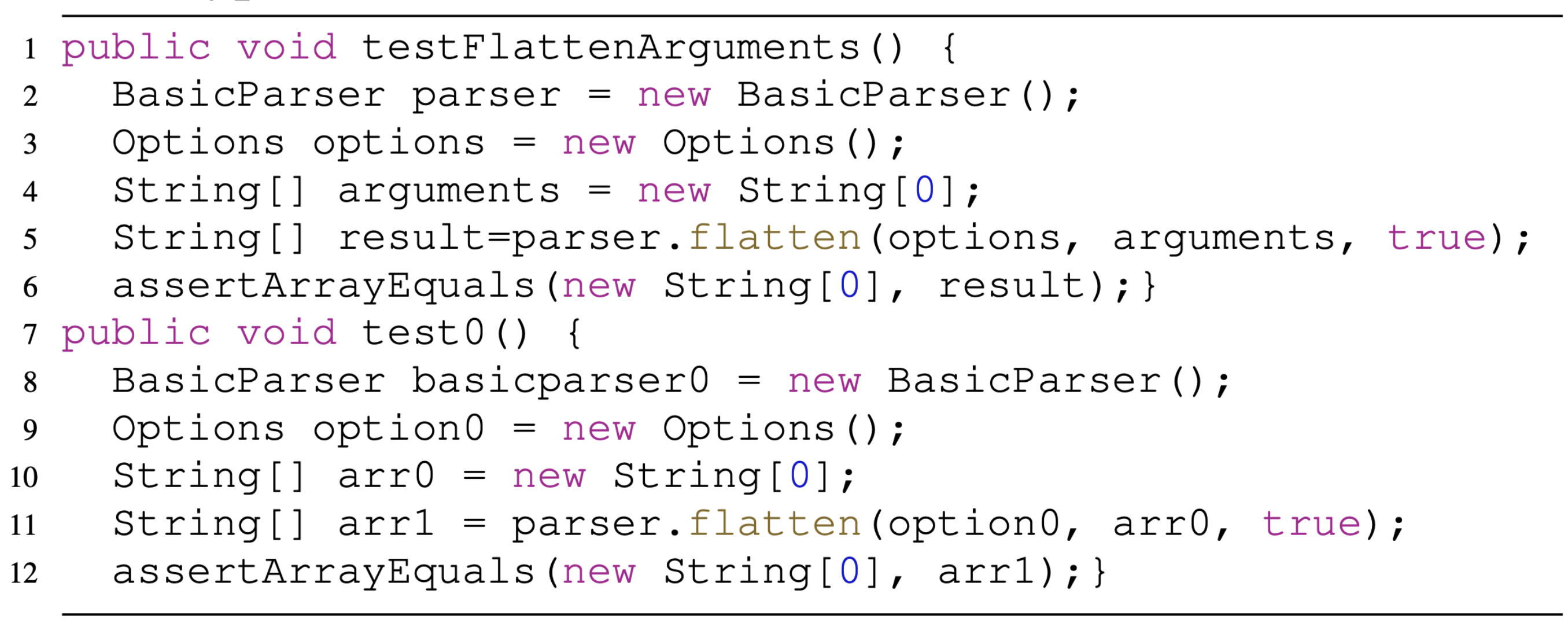}
		\centering
 \end{figure}
 \vspace{-5pt}
For instance, a variable of type \texttt{\small String} named \texttt{\small str} is not meaningful as it is not capturing any information about the stored data. In contrast, variable \texttt{\small BasicParser parser} is meaningful as the type name itself conveys meaning of the stored data. Additionally, a variable name's meaningfulness can depend on the context in which it is used. For example, in the case of \texttt{\small String[] arguments}, the variable name depends on the formal parameter names of the method \texttt{\small flatten(Options options, String[] arguments, boolean stopAtNonOption)} whose returned value is stored in the variable. Based on this premise, we categorize variables into two groups. 
Then, depending on the group, we determine whether to match with the data type name, assignment context, and formal parameter names, or simply the assignment context and formal parameter names. Finally, we use the Levenshtein distance to compute the closeness.

\begin{figure*}[t]
	\centering
	\begin{subfigure}{.215\textwidth}
		\centering
		\includegraphics[, ,width=\linewidth]{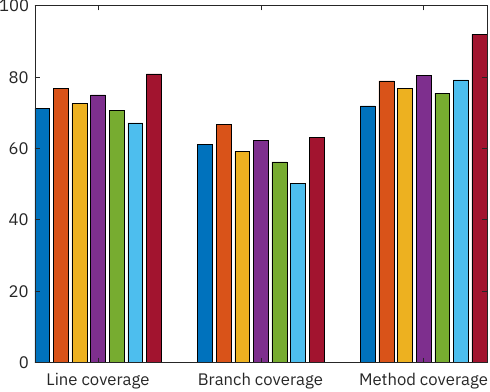}
		\centering
        \vskip -5pt
		\caption{\scriptsize Commons CLI}
	\label{fig:cli}
	\end{subfigure}%
 \hfill
	\begin{subfigure}{.215\textwidth}
		\centering
		\includegraphics[, ,width=\linewidth]{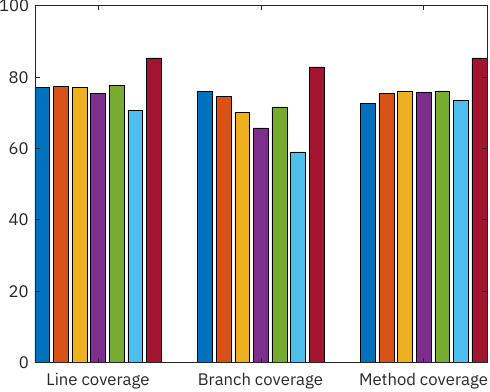}
        \vskip -5pt
		\caption{\scriptsize Commons Codec}
	\label{fig:codec}
	\end{subfigure}
  \hfill
    \begin{subfigure}{.215\textwidth}
		\centering
		\includegraphics[, ,width=\linewidth]{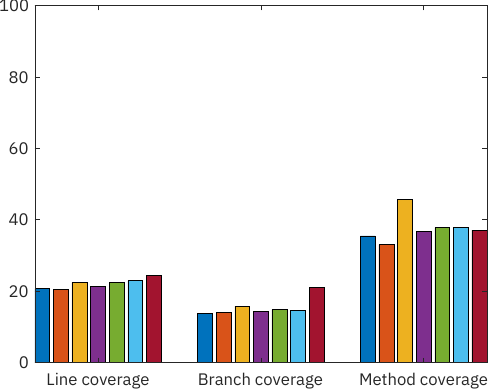}
		\centering
        \vskip -5pt
		\caption{\scriptsize Commons Compress}
	\label{fig:compress}
	\end{subfigure}%
 \hfill  
    \begin{subfigure}{.215\textwidth}
		\centering
		\includegraphics[, ,width=\linewidth]{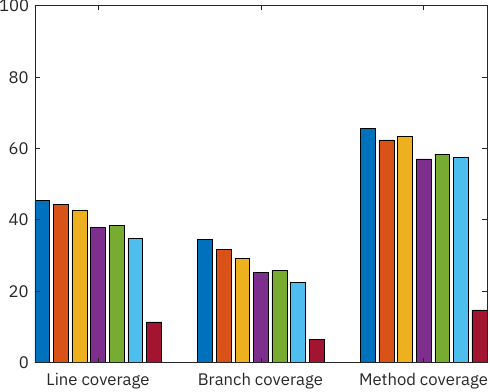}
		\centering
        \vskip -5pt
		\caption{\scriptsize Commons JXPath}
	\label{fig:jxpath}
	\end{subfigure}%
\vskip 3pt 


	\begin{subfigure}{.215\textwidth}
		\centering
		\includegraphics[, ,width=\linewidth]{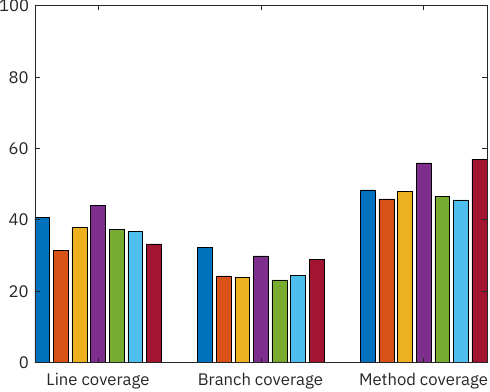}
		\centering
        \vskip -5pt
		\caption{\scriptsize CargoTracker}
	\label{fig:carogtracker}
	\end{subfigure}%
 \hfill
	\begin{subfigure}{.215\textwidth}
		\centering
		\includegraphics[, ,width=\linewidth]{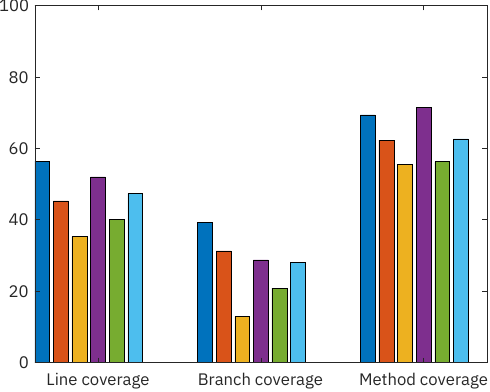}
        \vskip -5pt
		\caption{\scriptsize PetClinic}
	\label{fig:petclinic}
	\end{subfigure}
  \hfill
    \begin{subfigure}{.215\textwidth}
    \vspace{10pt}
		\centering
		\includegraphics[, ,width=\linewidth]{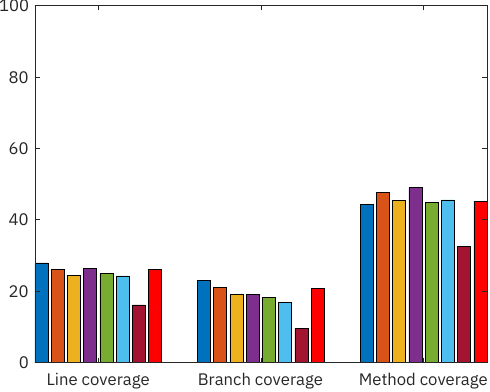}
		\centering
        \vskip -5pt
		\caption{\scriptsize DayTrader}
	\label{fig:daytrader}
	\end{subfigure}%
 \hfill
    \begin{subfigure}{.215\textwidth}
		\centering
		\includegraphics[, ,width=\linewidth]{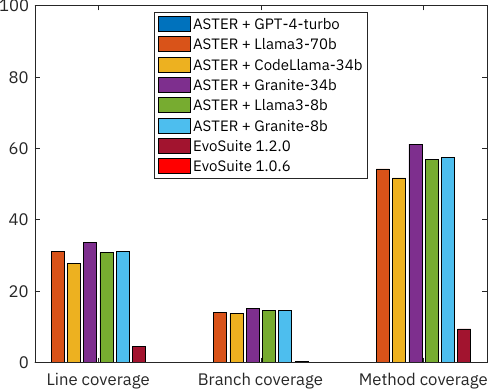}
		\centering
        \vskip -5pt
		\caption{\scriptsize App X}
	\label{fig:appx}
	\end{subfigure}%

 \vspace{7pt}
	 \caption{\small Line, branch, and method coverage achieved on Java SE and Java EE applications by \aster (configured with different LLMs) and EvoSuite (GPT-4 run excluded for App X for confidentiality reasons).}
  \label{fig:coverage-java}
  \vspace{-5pt}
\end{figure*}

\vskip 3pt
\noindent\textit{F. Experiment environment:}
The experiments were conducted on cloud VMs, each equipped with a 48-core Intel(R) Xeon(R) Platinum 8260 processor; the RAM ranged from 128 to 384 GB. We used the OpenAI API~\cite{openaiapi} to access the GPT models and an internal (proprietary) cloud service to access the other models.
We used v1.2.0 of EvoSuite and our updated version of CodaMosa. 
We performed three runs of test generation with each model for both Java dataset and the Python dataset. 
We use temperature = 0.2, which has been used for code generation tasks~\cite{ugare2024improving}, and for generating more predictable outputs, with max token length set to 1024 for reducing cost. 

\section{Evaluation Results}
\label{sec:evaluation}

\subsection{RQ1: Code Coverage}
\label{subsec:rq1coverage}


We collected code coverage in two steps, configuring \aster to run with each of the six models. In the first step, we ran \aster with its preprocessing and prompting components, followed by the postprocessing, i.e., to fix parsing/compilation/runtime errors and test failures. This step resulted in the initial generated test suite. Next, we performed coverage augmentation, asking the LLM to generate tests targeting lines of code not covered initially. We restricted augmentation to partially covered methods, omitting uncovered methods---the rationale is that methods that remain uncovered after the first step are unlikely to be covered during augmentation.

\textit{1. Java SE applications:}
\label{subsubsec:rq1a}
The top part of Figure~\ref{fig:coverage-java} presents coverage results for the Java SE applications in our dataset. The charts show, for each application, line, branch, and method coverage achieved by \aster configured with different LLMs and by EvoSuite. \aster is, in general, very competitive with EvoSuite---within a few percentage points, and even outperforming it, on different coverage metrics---with the best-performing model. For instance, for Commons CLI, \aster with Llama3-70B achieves only slightly lower line coverage (-4\%) and higher branch coverage (+3\%) than EvoSuite. Similarly, on Commons Codec, \aster with GPT-4 has slightly lower on line coverage (-7\%) and branch coverage (-7\%). On Commons Compress, both tools achieve much lower coverage, due to application complexity (Commons Compress provides an API for different file compression and archive formats), but there is not much difference between them. Finally, Commons JXPath highlights the case where \aster significantly outperforms EvoSuite, achieving 4x--5x more line and branch coverage with different models.
On examining the test cases for Commons JXPath, we found \aster's support for mocking to be one of the factors contributing to its superior performance; Figure~\ref{fig:naturlaness-examples}(b) illustrates an example test case. To understand the effect of our mock-generation approach, we conducted an ablation with Commons JXPath. We found that, on average, there is a (relative) loss of 13.7\%, 17.4\%, and 10.5\% in line, branch, and method coverage.
Overall, these results are positive and indicate that LLM-based test generation can match conventional test-generation tools on their one of their key strengths (code coverage), while also producing considerably more natural test cases (\S\ref{subsec:rq2naturalness}) that are preferred by developers (\S\ref{subsec:rq5survey}).

\begin{figure}
		\centering
		\includegraphics[, ,width=.56\linewidth]{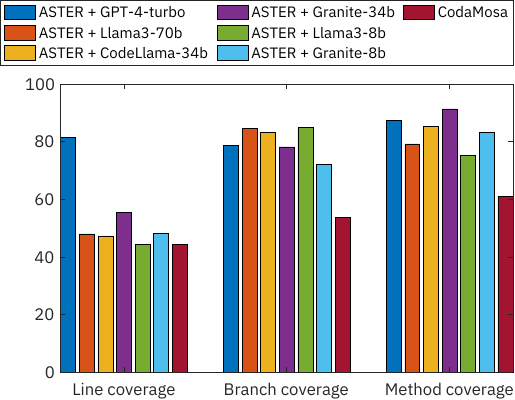}
		\centering
        \vspace{-7pt}
		\caption{\small Line, branch, and method coverage achieved on Python projects by \aster (with different LLMs) and CodaMosa.}
	\label{fig:coverage-python}
 \vspace{10pt}
 \end{figure}
\vspace{-5pt}
\begin{findingBox}{1}{
LLM-based test generation guided by static analysis is very competitive with EvoSuite in coverage achieved for Java SE projects, being slightly lower in some cases (-7\%) and considerably higher in other cases (4x--5x).
}\end{findingBox}
\vspace{-5pt}

Among the LLMs, we observe that all models perform roughly similar with respect to line and method coverage, but some differences become apparent on examining branch coverage. Another noteworthy result is that despite GPT-4's significantly larger size compared to the other models, its performance is not far superior to the smaller models; in some cases, the smaller models in fact perform better GPT-4 (Llama-70b: +1.2\%, CodeLlama-34b: +0.1\%, Granite-34b: -1.2\%, Llama-8b: -1.35\%, Granite-8b: -4.7\% w.r.t line coverage).

\textit{2. Java EE applications:}
\label{subsubsec:rq1b}
The bottom part of Figure~\ref{fig:coverage-java} presents coverage results for Java EE applications. The result for CargoTracker is similar to that for Java SE applications, with \aster being competitive with EvoSuite and, in some instances, performing better than it. For PetClinic, EvoSuite could not be run because PetClinic requires Java~17, which is not supported by EvoSuite. This highlights the benefit of LLM-based test generation that it can inherently support more language versions than conventional tools.
On DayTrader and App X, \aster consistently achieves higher scores on all coverage metrics, with the difference for App X being especially significant---7x more line coverage and 84x more branch coverage with the best-performing model, Granite-34B. We investigated the reasons for this difference and found \aster's mocking capability and handling of multiple frameworks (stemming from the LLM's inherent knowledge) as the two primary reasons.
An older release of EvoSuite (1.0.6) supports Java~EE~features but works with Java~8 only. Among the Java~EE~applications in our dataset, DayTrader is the only one that can be built with Java~8. Thus, we applied EvoSuite~1.0.6 to DayTrader, which yielded much better results, raising EvoSuite's performance to a similar level as \aster, as shown in Figure~\ref{fig:coverage-java}(g).
\vspace{-5pt}
\begin{findingBox}{2}{For Java EE projects, \aster significantly outperforms EvoSuite for most applications (10.6\%-26.4\% coverage) and is capable of generating test cases for applications where existing approaches fail to do so.
}
\end{findingBox}
\vspace{-5pt}

 Our findings suggest that smaller models (8b, 34b) can match or even outperform larger models such as Llama3-70b or GPT-4. This is promising because the major drawback of LLM-based test generation is the associated cost. Because this process involves thousands of LLM calls with token-based billing for accessing models through their APIs, the cost can be significantly high. 
 Smaller models also require less computational power, resulting in smaller carbon footprint. Additionally, these models can operate on workstations or on-premise devices, making them ideal for enterprise use cases where restrictions prevent code and data from being transmitted outside the organization.
\vspace{-5pt}
\begin{findingBox}{3}{Smaller models (Granite-34b and Llama-3-8b) demonstrate competitive performance, with only 0.1\%, 6.3\%, and 2.7\% loss in line, branch, and method coverage, compared to larger models (Llama-70b and GPT-4).
}\end{findingBox}
\vspace{-5pt}

\textit{3. Python applications:}
Figure~\ref{fig:coverage-python} presents coverage results for Python: it shows line, branch, and method coverage achieved by \aster configured with the six LLMs compared with CodaMosa over the Python dataset. \aster performs considerably better than CodaMosa (CodaMosa: 44\%, 53.7\%, and 61.2\% line, branch, and method coverage, \aster + GPT-4: 78\% line coverage, 77.2\% branch coverage, and 86.7\% method coverage).
\aster with Granite-34B performed even better in method coverage, reaching 89.9\%.
Notably, the smaller models performed well too. With Granite-8b, \aster achieved method, branch, and line coverage of 83\%, 71.9\%, and 48.3\%, respectively. 
These results validate the effectiveness of \aster's LLM-driven approach, demonstrating a clear advantage over conventional techniques.



\begin{findingBox}{4}{\aster generates Python tests with higher coverage (+9.8\%, +26.5\%, and +22.5\%) for all the models compared to CodaMosa.
}\end{findingBox}







\subsection{RQ2: Developer Survey}

\label{subsec:rq5survey}
\begin{figure*}
    \centering
\begin{subfigure}{.5\textwidth}
		\centering
		\includegraphics[, ,width=0.91\linewidth]{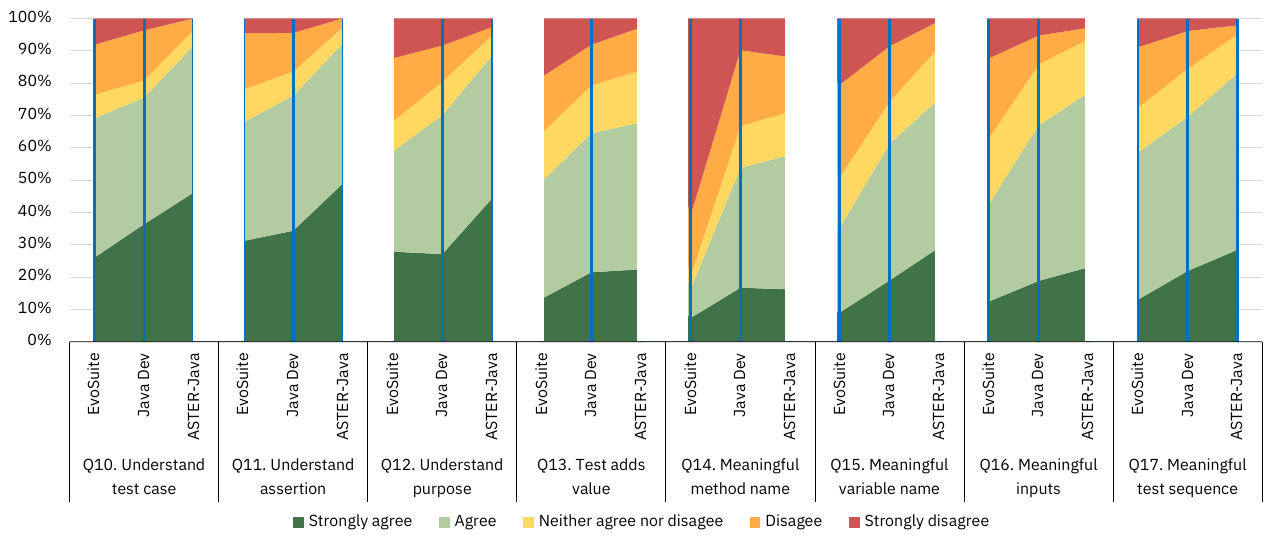}
		\centering
	\label{fig:surveyjava}
	\end{subfigure}%
 \hfill
	\begin{subfigure}{.5\textwidth}
		\centering
		\includegraphics[, ,width=0.91\linewidth]{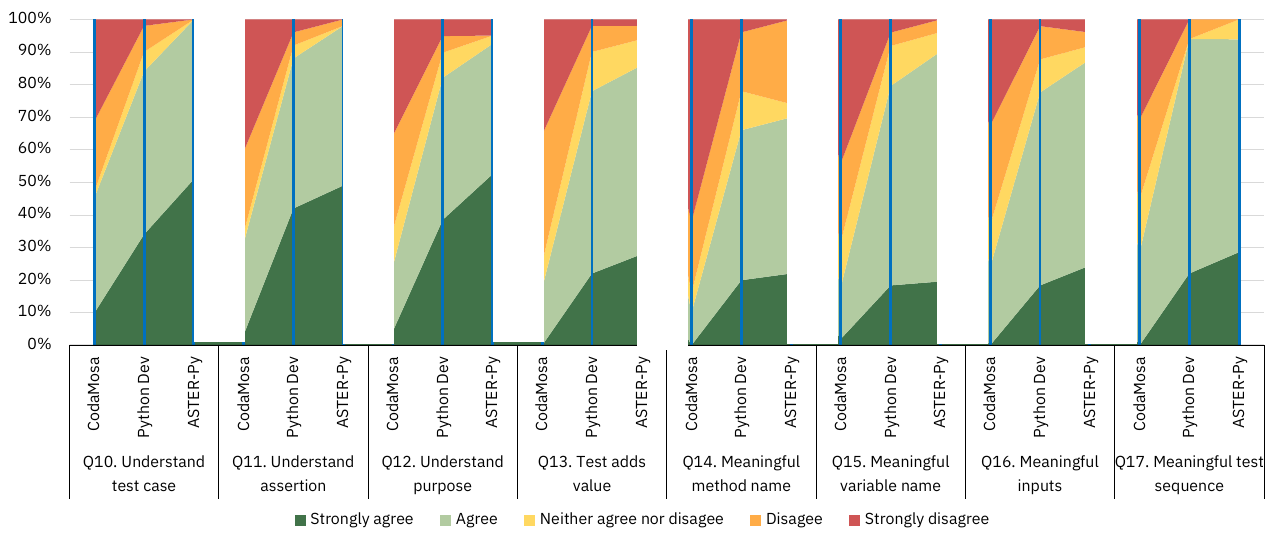}
		\centering
	\label{fig:surveypython}
	\end{subfigure}%
    \vspace{-7pt}
    \caption{\small Survey responses on test quality (Q10--Q17) for Java test cases (left) and Python test cases (right).}
    \vspace{-7pt}
    \label{fig:surveyresult}
\end{figure*}
\begin{figure}[!htp]
    \centering
    \includegraphics[width=0.74\linewidth]{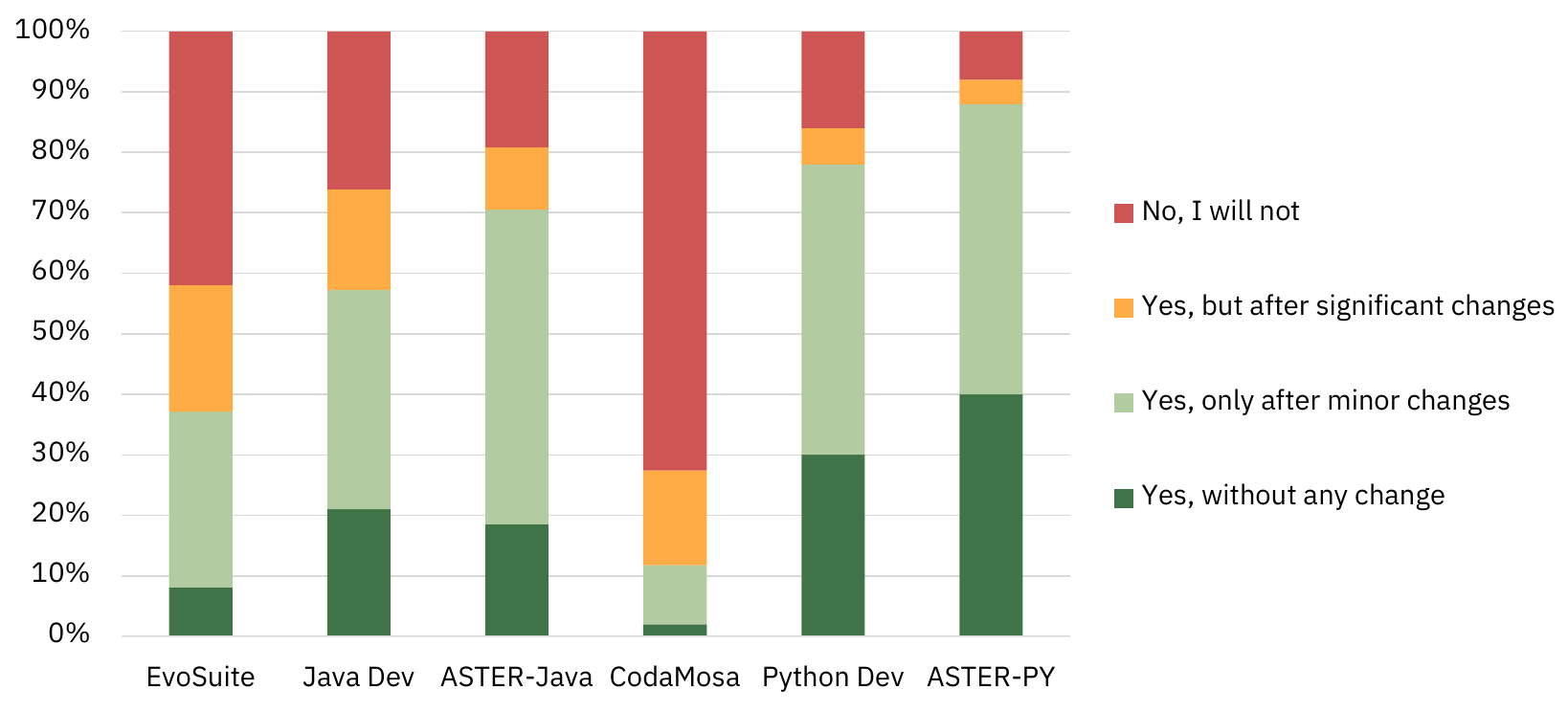}
    \vspace{-7pt}
    \caption{\small Survey response on Q18 (whether developers would add such test cases to their test suites).}
    \vspace{7pt}
    \label{fig:surveyq18}
\end{figure}
To understand developer perspectives on comprehensibility and usability of \aster-generated tests compared to EvoSuite (or CodaMosa)-generated and developer-written tests, we conducted an anonymous online survey in an industry setting. The survey consists of a set of background questions, followed by a series of focal methods together with two test cases for each method, and a set of questions for each focal method and its test pair. Figure~\ref{tb:survey} shows the survey questions.
\begin{table}[t]
  \centering
  \setlength{\tabcolsep}{3pt}
  \scriptsize
  \caption{Two groups of survey questions.}  
  \resizebox{0.7\linewidth}{!}{
    \begin{tabular}{p{8pt}p{6.5cm}p{0.5cm}}
    \toprule
    \multicolumn{1}{c}{\bf Type} & \multicolumn{1}{c}{\bf Question} & \multicolumn{1}{c}{\bf Format} \\
    \midrule
    \multirow{9}{*}{\begin{sideways}Background\end{sideways}} & {Q1.} Current Professional Role & Open    \\
          & {Q2.} Years of experience in software engineering (incl. education) & MCQ    \\
          & {Q3.} Years of experience in industry & MCQ     \\
          & {Q4.} Level of expertise in Java & MCQ     \\
          & {Q5.} Level of expertise in Python & MCQ     \\
          & {Q6.} Prior experience with automated test generation & MCQ     \\
          & {Q7.} Prior experience with automated test generation (if yes, list the tools used in the past)& MCQ, Open   \\
          & {Q8.} How often do you write unit tests for your code? & MCQ     \\
          & {Q9.} On average, how long do you spend writing good unit tests for a single method? & MCQ     \\
    \midrule
    \multirow{10}{*}{\begin{sideways}Test Quality\end{sideways}} & {Q10.} I understand what this test case is doing & Likert   \\
          & {Q11.} I understand what the assertions in this test are checking & Likert   \\
          & {Q12.} I can describe the purpose of this test case & Likert   \\
          & {Q13.} This test case adds value & Likert   \\
          & {Q14.} The test name correctly describes what is being tested & Likert   \\
          & {Q15. }The names of variables in the test case are meaningful & Likert   \\
          & {Q16.} The input values (e.g., string literals, integer constants), if any, used in the test case are meaningful & Likert   \\
          & {Q17. }The test sequence (i.e., the sequence of method calls in the test case) makes sense & Likert   \\
          & {Q18.} Would you add this test case to your unit test bucket? & MCQ     \\
          & {Q19.} Please provide any descriptive feedback. & Open    \\
    \bottomrule
    \end{tabular}%
    }
  \label{tb:survey}
\end{table}%

We used two Java (Commons CLI and Commons JXPath) and three Python applications (Ansible~\cite{ansible}, Tornado Web~\cite{tornadoweb}, and Flutes~\cite{flutes}) for the study. We designed the survey to present, for each focal method, a pair of test cases, where the pair could be (\aster, EvoSuite/CodaMosa), (\aster, developer), or (developer, EvoSuite/CodaMosa) tests---without indicating the source of a test case. To select focal methods, first we mapped each test case, from the three sources, to its focal method, and removed focal methods that did not have at least one source-pair test.
Then, we randomly selected focal methods 
and a pair of tests for each method, while ensuring that the participant would see an equal number of tests from the three sources. In total, the survey had 9 focal methods and 18 test cases, with 6 tests from each test source. The survey presents the participant with the focal method's body, a brief task description, and a GitHub repository URL (using which the participant could, if needed, browse the repository content). 

The survey received 161~responses, with participants coming various roles, such as software developer, QA engineer, principal solution architect, research scientist, etc. In terms of experience,
69.7\% of the participants have $>$10 years of experience, with 22.2\% exceeding 25 years. Most participants have very strong industry experience, with 48.1\% of the participants being in industry for 10+ years.
Participants also have high level of proficiency in Java and Python, with 71\% and 62\% reporting professional or experienced levels, respectively. A few participants said they have used code assistants as aids in test generation. Additionally, all participants agreed that writing test cases is tedious and 90.4\% said that writing a good unit test takes more than five minutes.
For Q10--Q18, participants provided responses on a five-point scale, ranging from ``strongly agree'' to ``strongly disagree''. The results, shown in Fig.~\ref{fig:surveyresult}, indicate a strong preference for \aster-generated test cases over EvoSuite and CodaMosa for all aspects evaluated. For example, on Q10, 91.6\% of the respondents agreed or strongly agreed that they understood the \aster-generated Java test cases, which is considerably higher than their ratings for EvoSuite tests (67.7\%) and the developer-written tests (69.1\%). On this question, the difference for the Python tests is much higher, with 100\% positive responses on \aster-generated tests, compared with 84.0\% for developer-written tests and only 45.1\% for the CodaMosa tests. Overall, on all the questions, the participants rated \aster-generated tests higher than EvoSuite/CodaMosa-generated tests and even developer-written tests in most cases.
Finally, on Q18 (Fig.~\ref{fig:surveyq18}), participants indicated a much stronger preference for incorporating \aster-generated tests into their test suites with minor or no changes---70\% for Java and 88\% for Python---compared to EvoSuite and CodaMosa tests. On this question too, \aster-generated tests score higher than developer-written tests.
\vspace{-5pt}
\begin{findingBox}{5}{Developers prefer \aster-generated tests over EvoSuite and CodaMosa tests, with over 70\% also willing to add such tests with minor or no changes to their test buckets. \aster-generated tests are also favored slightly over developer-written tests in these characteristics.  
}\end{findingBox}
\vspace{-5pt}

\subsection{RQ3: Test Naturalness}
\label{subsec:rq2naturalness}
\begin{figure*}
    \centering
    \includegraphics[width=0.91\textwidth]{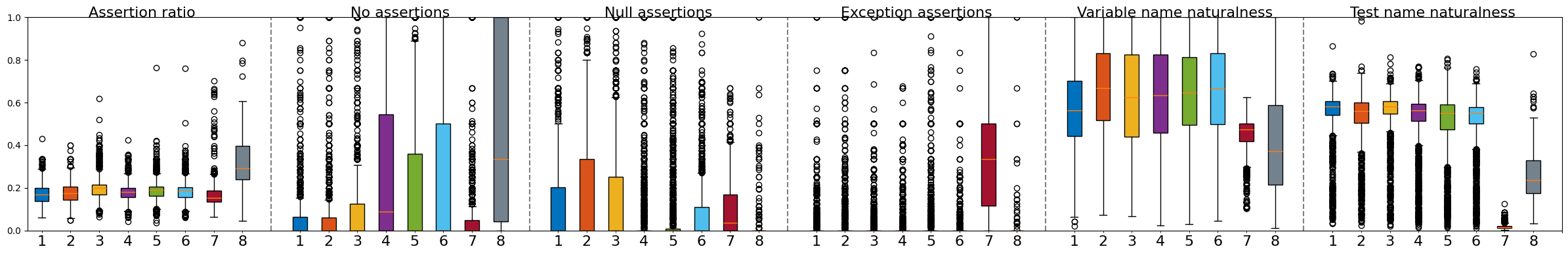}
    \vspace{-7pt}
    \caption{\small Naturalness results for Java SE and EE applications (1: GPT-4, 2: Llama3-70b, 3: CodeLlama-34b, 4: Granite-34b, 5: Llama-8b, 6: Granite-8b, 7: EvoSuite, 8: Developer).}
    \vspace{-8pt}
    \label{fig:javanat}
\end{figure*}
\begin{figure}
    \centering
    \includegraphics[width=0.86\columnwidth]{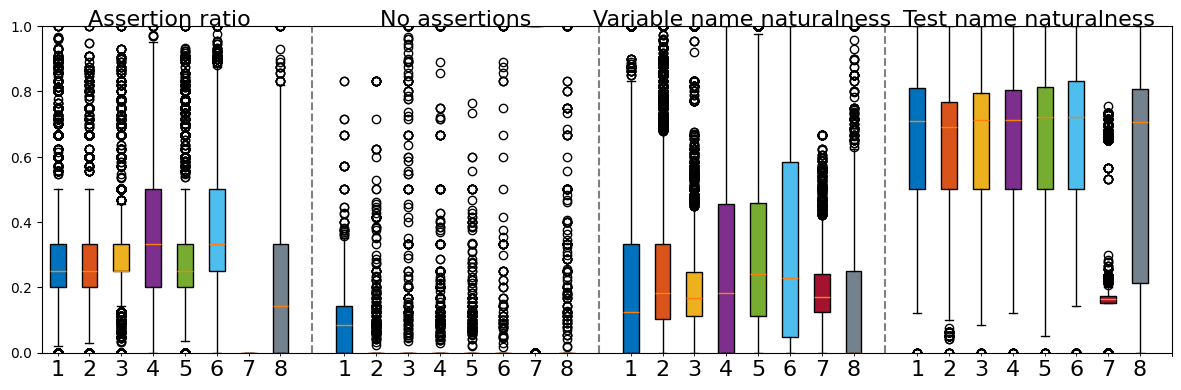}
    \vspace{-7pt}
    \caption{\small Naturalness results for Python (legends are same as Figure~\ref{fig:javanat} expect 7 represents CodaMosa).}
    \vspace{7pt}
    \label{fig:pynat}
\end{figure}
We ran our test naturalness analyzer on all the Java and Python applications and reported results on assertion ratio, percentage of tests with no assertion, null assertions, exception-related assertions, and also reported the score for variable and test name naturalness (\cref{fig:javanat} for Java and \cref{fig:pynat} for Python).

\textit{1. Java applications:}
\label{subsubsec:rq2a}
We discovered that both EvoSuite-generated and LLM-generated tests exhibit test smells. Interestingly, LLM-generated tests tend to have more cases (avg. 25\%) without assertions compared to those generated by EvoSuite. This opens up an intriguing avenue for enhancing LLM-generated tests to reduce smells while maintaining coverage. Notably, \aster-generated test cases contain significantly fewer exception-related tests than those generated by EvoSuite (avg. 38\%). In terms of readability metrics, \aster-generated test cases outperform those generated by EvoSuite. \aster-generated tests have test names that are $\sim$15\% and $\sim$23\% more natural compared to EvoSuite and developer-written test cases.  However, there is still a significant room for improvement and we believe a finetuned model for test generation can reduce the gap. Additionally, our automated approach has certain limitations, such as not capturing natural language descriptions in test names effectively. For example, in \texttt{\footnotesize testPrintGainHTML_PositiveGain}, the term \texttt{\footnotesize PositiveGain} is not directly reflected in the code body. 

\textit{2. Python applications:}
\label{subsubsec:rq2b}
In our study using CodaMOSA's replication package, we observed that the generated test cases consistently lacked assertions. It's important to note that while Pynguin, a baseline tool for CodaMOSA, has the capability to generate assertions, this feature was disabled in the replication package we used. As discussed in the CodaMOSA paper~\cite{codamosa}, their focus is on achieving high code coverage, and do not consider assertions. In contrast, \aster places a strong emphasis on assertion generation, producing test cases with assertions in the vast majority of cases (99\%).
Only GPT-generated test cases have 8\% of test cases with no assertions. None of the developer-written, \aster-generated, or CodaMosa-generated tests (no assertion) contain null assertions or exception assertions. 
In terms of variable and test name naturalness, \aster achieved scores very similar to developer-written test cases, while the naturalness score of CodaMosa was 55\% lower compared to developer-written and \aster-generated tests.
\vspace{-5pt}
\begin{findingBox}{6}{\aster-generated test cases have more meaningful test and variable names compared to EvoSuite and CodaMosa tests. 
}\end{findingBox}
\vspace{-5pt}





\section{Lessons Learned}
\label{sec:discussion}


\textit{Supporting more PLs and frameworks.}
LLMs, when combined with lightweight program analysis, offer versatile support for multiple PLs and frameworks. The \aster pipeline demonstrates this adaptability, often matching or surpassing existing methods in test generation across various languages. To extend support, developers can utilize tools like Tree-sitter~\cite{treesitter}, CodeQL~\cite{codeql}, and WALA~\cite{wala}, addressing the crucial need for multi-language and multi-framework support in both legacy and continuous development enterprise contexts.

\textit{Generating more natural test cases.} Developers prefer naturalness in test cases, including intuitive test names, variable names, and high-quality assertions. LLMs, trained on developer-written code, inherently produce more natural output compared to existing ATGs. While LLMs can make mistakes, combining them with program analysis yields high-quality test cases in terms of both coverage and naturalness. Potential research directions in this area include enhancing the naturalness of conventionally generated tests and incorporating naturalness as a pretraining objective for LLMs. 



\textit{Affordability and privacy.} LLM-based test generation faces affordability and privacy challenges in enterprise settings. Our study shows smaller models such as Granite-34b and Llama3-8b perform competitively, suggesting cost-effective alternatives to GPT-4. Privacy concerns necessitate on-premise solutions, with developers preferring models hosted internally or on local workstations. Future research should focus on developing efficient, quantized models specific to test generation, and reducing LLM calls through deterministic approaches. 

\section{Related Work}
\label{sec:related}

Compared to the previous work~\cite{fraser2011evosuite, arcuri2019restful, pacheco2007feedback, pacheco2007randoop, lukasczyk2022pynguin, tillmann2008pex, sen2005cute,godefroid2005dart, tzoref2022tackletest} on conventional test-generation approaches, we leverage LLMs to generate more natural unit tests while supporting multiple PLs.
Recently, there have been several attempts at employing LLMs for test generation~\cite{wang2024software, tufano2020unit, vikram2023can, bareiss2022code,ryan2024code, schafer2023empirical, tufano2022generating}. 
AthenaTest~\cite{tufano2020unit} developed a fine-tuning pipeline for test generation. Vikram~\etal~\cite{vikram2023can} leveraged LLMs to support property-based testing. Ryan~\etal~\cite{ryan2024code} adopted symbolic analysis to boost coverage and Bareiß~\etal~\cite{bareiss2022code} incorporated few-shot examples to guide test generation. TestPilot~\cite{schafer2023empirical} synthesizes unit test cases for JavaScript and TypeScript by gathering usage documentation and API functions.
 ChatTester~\cite{yuan2024evaluating} and ChatUnitTest~\cite{chen2024chatunitest} used the GPT model to generate test cases given the information related to the class under test and have shown that it performs competitively. However, this approach is specifically tied to GPT, and in many enterprise use cases, this approach may not work. Recent Pizzorno and Berger~\cite{pizzorno2024coverup} (arxiv), use LLM as a coverage-augmenting approach starting with tests generated by CodaMosa. Compared to that work, \aster supports more PLs, generates tests from scratch, creates more natural tests than CodaMosa, which the prior work is dependent on, and generates tests with higher coverage than CodaMosa (prior work: line: +7.3\%, branch: +10.3\%, \aster: line: +37.1\%, branch: +24.9\% with GPT-4, on which the prior work was evaluated). 

Compared to the studies done in different industry settings~\cite{meta2023testingchallenges, meta2024, meta2024observationbased, schafer2023empirical}, our work provides several new findings and also confirms various important lessons learned in the prior work. For instance, the acceptance rate of the LLM-generated tests reported by the work done at Meta~\cite{meta2024} and our work are very similar (73\% at Meta and $>$70\% in our case), which shows that developers overall prefer such test cases. Moreover, both studies identify test naturalness as a key factor in developers' acceptance criteria. 
Another common finding of these studies is that using static and dynamic analysis help increase the performance of the LLMs. Our work provides several new findings, related to generation of tests with API mocking, multi-language test generation, performance of models of different sizes/families, and a rigorous study of test naturalness including a developer survey.



\section{Threats to Validity}
\label{threats}


To address threat related to generalizability of \aster, we extended support for (a) multiple PLs and (b) multiple models with varying sizes, modalities, and families. Another potential threat is the limited number of evaluation runs. While previous studies have performed $>10$ runs, the substantial cost associated with running evaluations across several applications and six LLMs led us to limit the runs. 
Another potential threat is the automated naturalness evaluation; to mitigate that, we conducted a survey of professional developers, and found that the findings are very similar.

\section{Summary and Future Works}
\label{conclusion}

In this paper, we presented \aster, a multi-language test-generation tool that leverages LLMs guided by lightweight static analysis to generate natural and effective unit test cases for Java and Python. Through its preprocessing component, \aster ensures that LLM prompts have adequate context required for generating unit tests for a focal method. \aster's postprocessing component performs iterative test repair and coverage augmentation. Our extensive evaluation, with six LLMs on a dataset of Java SE, Java EE, and Python application, showed that \aster is competitive with state-of-the-art tools in coverage achieved on Java SE application, and outperforms them significantly on Java EE and Python applications, while also producing considerably more natural tests than those tools. Our developer survey, with over 160 participants, highlighted the naturalness characteristics of \aster-generated tests and their usability for building automated test suites. 
Future research directions include extending \aster to other PLs and levels of testing (e.g., integration testing), creating fine-tuned models for testing to reduce the cost of LLM interactions, and exploring techniques for improving fault-detection ability of the generated tests. 
\section*{Acknowledgment}
We thank Xuan Liu, Maja Vuković, Nicholas Fuller, Angel Montesdeoca, Lisa Waugh, Jayant Talekar, and all the members of IBM watsonx Code Assistant team for their help with this research. We also thank the anonymous reviewers for their comments, which helped make this work stronger.

\IEEEtriggeratref{59}
\bibliographystyle{IEEEtran}
\bibliography{refs}

\begin{thebibliography}{10}
\providecommand{\url}[1]{#1}
\csname url@samestyle\endcsname
\providecommand{\newblock}{\relax}
\providecommand{\bibinfo}[2]{#2}
\providecommand{\BIBentrySTDinterwordspacing}{\spaceskip=0pt\relax}
\providecommand{\BIBentryALTinterwordstretchfactor}{4}
\providecommand{\BIBentryALTinterwordspacing}{\spaceskip=\fontdimen2\font plus
\BIBentryALTinterwordstretchfactor\fontdimen3\font minus \fontdimen4\font\relax}
\providecommand{\BIBforeignlanguage}[2]{{%
\expandafter\ifx\csname l@#1\endcsname\relax
\typeout{** WARNING: IEEEtran.bst: No hyphenation pattern has been}%
\typeout{** loaded for the language `#1'. Using the pattern for}%
\typeout{** the default language instead.}%
\else
\language=\csname l@#1\endcsname
\fi
#2}}
\providecommand{\BIBdecl}{\relax}
\BIBdecl

\bibitem{visser:2004}
W.~Visser, C.~S. Pasareanu, and S.~Khurshid, ``Test input generation with java pathfinder,'' in \emph{Proceedings of the 2004 ACM SIGSOFT International Symposium on Software Testing and Analysis}, 2004, p. 97–107.

\bibitem{cadar:2008}
C.~Cadar, D.~Dunbar, and D.~Engler, ``Klee: Unassisted and automatic generation of high-coverage tests for complex systems programs,'' in \emph{Proceedings of the 8th USENIX Conference on Operating Systems Design and Implementation}, 2008, p. 209–224.

\bibitem{pasareanu:2010}
C.~S. P\u{a}s\u{a}reanu and N.~Rungta, ``Symbolic pathfinder: Symbolic execution of java bytecode,'' in \emph{Proceedings of the IEEE/ACM International Conference on Automated Software Engineering}, 2010, p. 179–180.

\bibitem{godefroid2005dart}
P.~Godefroid, N.~Klarlund, and K.~Sen, ``Dart: Directed automated random testing,'' in \emph{Proceedings of the 2005 ACM SIGPLAN conference on Programming language design and implementation}, 2005, pp. 213--223.

\bibitem{sen2005cute}
K.~Sen, D.~Marinov, and G.~Agha, ``Cute: A concolic unit testing engine for c,'' \emph{ACM SIGSOFT Software Engineering Notes}, vol.~30, no.~5, pp. 263--272, 2005.

\bibitem{xie:2005}
T.~Xie, D.~Marinov, W.~Schulte, and D.~Notkin, ``Symstra: A framework for generating object-oriented unit tests using symbolic execution,'' in \emph{Proceedings of the 11th International Conference on Tools and Algorithms for the Construction and Analysis of Systems}, 2005, p. 365–381.

\bibitem{tillmann2008pex}
N.~Tillmann and J.~De~Halleux, ``Pex--white box test generation for. net,'' in \emph{International conference on tests and proofs}.\hskip 1em plus 0.5em minus 0.4em\relax Springer, 2008, pp. 134--153.

\bibitem{tonella:2008}
P.~Tonella, ``Evolutionary testing of classes,'' in \emph{Proceedings of the 2004 ACM SIGSOFT International Symposium on Software Testing and Analysis}, 2004, p. 119–128.

\bibitem{harman:2010:tse}
M.~Harman and P.~McMinn, ``A theoretical and empirical study of search-based testing: Local, global, and hybrid search,'' \emph{IEEE Transactions on Software Engineering}, vol.~36, no.~2, pp. 226--247, 2010.

\bibitem{mcminn:2004}
P.~McMinn, ``Search-based software test data generation: A survey: Research articles,'' \emph{Softw. Test. Verif. Reliab.}, vol.~14, no.~2, pp. 105--156, Jun. 2004.

\bibitem{fraser2011evosuite}
G.~Fraser and A.~Arcuri, ``{EvoSuite}: Automatic test suite generation for object-oriented software,'' in \emph{Proceedings of the 19th ACM SIGSOFT symposium and the 13th European conference on Foundations of software engineering}, 2011, pp. 416--419.

\bibitem{lin:2021}
Y.~Lin, Y.~S. Ong, J.~Sun, G.~Fraser, and J.~S. Dong, ``Graph-based seed object synthesis for search-based unit testing,'' in \emph{Proceedings of the 29th ACM Joint Meeting on European Software Engineering Conference and Symposium on the Foundations of Software Engineering}, 2021, p. 1068–1080.

\bibitem{lukasczyk2022pynguin}
S.~Lukasczyk and G.~Fraser, ``Pynguin: Automated unit test generation for python,'' in \emph{Proceedings of the ACM/IEEE 44th International Conference on Software Engineering: Companion Proceedings}, 2022, pp. 168--172.

\bibitem{pacheco2007feedback}
C.~Pacheco, S.~K. Lahiri, M.~D. Ernst, and T.~Ball, ``Feedback-directed random test generation,'' in \emph{29th International Conference on Software Engineering (ICSE'07)}.\hskip 1em plus 0.5em minus 0.4em\relax IEEE, 2007, pp. 75--84.

\bibitem{ciupa:2008:icse}
I.~Ciupa, A.~Leitner, M.~Oriol, and B.~Meyer, ``{ARTOO}: Adaptive random testing for object-oriented software,'' in \emph{Proceedings of the 30th International Conference on Software Engineering}, 2008, p. 71–80.

\bibitem{chen:2010:jss}
T.~Y. Chen, F.-C. Kuo, R.~G. Merkel, and T.~H. Tse, ``Adaptive random testing: The {ART} of test case diversity,'' \emph{J. Syst. Softw.}, vol.~83, no.~1, p. 60–66, Jan. 2010.

\bibitem{lin:2009:ase}
Y.~Lin, X.~Tang, Y.~Chen, and J.~Zhao, ``A divergence-oriented approach to adaptive random testing of java programs,'' in \emph{2009 IEEE/ACM International Conference on Automated Software Engineering}, 2009, pp. 221--232.

\bibitem{arcuri:2011:issta}
A.~Arcuri and L.~Briand, ``Adaptive random testing: An illusion of effectiveness?'' in \emph{Proceedings of the 2011 International Symposium on Software Testing and Analysis}, 2011, p. 265–275.

\bibitem{lukasczyk:2023:emse}
S.~Lukasczyk, F.~Kroi{\ss}, and G.~Fraser, ``An empirical study of automated unit test generation for python,'' \emph{Empirical Software Engineering}, vol.~28, no.~2, 2023.

\bibitem{fraser:2015}
G.~Fraser, M.~Staats, P.~McMinn, A.~Arcuri, and F.~Padberg, ``Does automated unit test generation really help software testers? a controlled empirical study,'' \emph{ACM Trans. Softw. Eng. Methodol.}, Sep. 2015.

\bibitem{panichella:2020:testsmells}
A.~Panichella, S.~Panichella, G.~Fraser, A.~A. Sawant, and V.~J. Hellendoorn, ``Revisiting test smells in automatically generated tests: Limitations, pitfalls, and opportunities,'' in \emph{2020 IEEE International Conference on Software Maintenance and Evolution (ICSME)}, 2020, pp. 523--533.

\bibitem{evosuite}
\BIBentryALTinterwordspacing
``{EvoSuite: Automatic Test Suite Generation for Java},'' 2024. [Online]. Available: \url{https://www.evosuite.org/}
\BIBentrySTDinterwordspacing

\bibitem{codamosa}
\BIBentryALTinterwordspacing
``{CodaMosa},'' 2024. [Online]. Available: \url{https://github.com/microsoft/codamosa}
\BIBentrySTDinterwordspacing

\bibitem{lemieux2023codamosa}
C.~Lemieux, J.~P. Inala, S.~K. Lahiri, and S.~Sen, ``Codamosa: Escaping coverage plateaus in test generation with pre-trained large language models,'' in \emph{2023 IEEE/ACM 45th International Conference on Software Engineering (ICSE)}.\hskip 1em plus 0.5em minus 0.4em\relax IEEE, 2023, pp. 919--931.

\bibitem{meta2024}
N.~Alshahwan, J.~Chheda, A.~Finogenova, B.~Gokkaya, M.~Harman, I.~Harper, A.~Marginean, S.~Sengupta, and E.~Wang, ``Automated unit test improvement using large language models at meta,'' in \emph{Companion Proceedings of the 32nd ACM International Conference on the Foundations of Software Engineering}, 2024, pp. 185--196.

\bibitem{ibm:wca4eja}
\BIBentryALTinterwordspacing
IBM, ``{watsonx Code Assistant for Enterprise Java Applications},'' 2024. [Online]. Available: \url{https://www.ibm.com/products/watsonx-code-assistant-for-enterprise-java-applications}
\BIBentrySTDinterwordspacing

\bibitem{aster:artifact}
\BIBentryALTinterwordspacing
``{ASTER Artifact},'' 2024. [Online]. Available: \url{https://github.com/aster-test-generation/aster}
\BIBentrySTDinterwordspacing

\bibitem{commonscli}
\BIBentryALTinterwordspacing
``{Apache Commons CLI},'' 2024. [Online]. Available: \url{https://github.com/apache/commons-cli}
\BIBentrySTDinterwordspacing

\bibitem{commonsjxpath}
\BIBentryALTinterwordspacing
``{Apache Commons JXPath},'' 2024. [Online]. Available: \url{https://github.com/apache/commons-jxpath}
\BIBentrySTDinterwordspacing

\bibitem{pylint}
pylint dev, ``{Pylint},'' accessed: 2024-07-27.

\bibitem{pytest}
N.~Batchelder, ``{Coverage.py: Code coverage measurement for Python},'' accessed: 2024-07-27.

\bibitem{coverage}
------, ``{Coverage.py: Code coverage measurement for Python},'' \url{https://coverage.readthedocs.io/}, accessed: 2024-07-27.

\bibitem{openai:deprecations}
\BIBentryALTinterwordspacing
OpenAI, ``Openai deprecations,'' 2024. [Online]. Available: \url{https://platform.openai.com/docs/deprecations}
\BIBentrySTDinterwordspacing

\bibitem{openaimodels}
\BIBentryALTinterwordspacing
------, ``Openai models,'' 2024. [Online]. Available: \url{https://platform.openai.com/docs/models/gpt-base}
\BIBentrySTDinterwordspacing

\bibitem{llama3models}
\BIBentryALTinterwordspacing
Meta, ``Meta llama 3,'' 2024. [Online]. Available: \url{https://huggingface.co/collections/meta-llama/meta-llama-3-66214712577ca38149ebb2b6}
\BIBentrySTDinterwordspacing

\bibitem{granitecodemodels}
\BIBentryALTinterwordspacing
IBM, ``Granite code models,'' 2024. [Online]. Available: \url{https://huggingface.co/collections/ibm-granite/granite-code-models-6624c5cec322e4c148c8b330}
\BIBentrySTDinterwordspacing

\bibitem{commonscodec}
\BIBentryALTinterwordspacing
``{Apache Commons Codec},'' 2024. [Online]. Available: \url{https://github.com/apache/commons-codec}
\BIBentrySTDinterwordspacing

\bibitem{commonscompress}
\BIBentryALTinterwordspacing
``{Apache Commons Compress},'' 2024. [Online]. Available: \url{https://github.com/apache/commons-compress}
\BIBentrySTDinterwordspacing

\bibitem{cargotracker}
\BIBentryALTinterwordspacing
``{Eclipse Cargo Tracker},'' 2024. [Online]. Available: \url{https://github.com/eclipse-ee4j/cargotracker}
\BIBentrySTDinterwordspacing

\bibitem{daytrader}
\BIBentryALTinterwordspacing
``{DayTrader8 Sample},'' 2024. [Online]. Available: \url{https://github.com/OpenLiberty/sample.daytrader8}
\BIBentrySTDinterwordspacing

\bibitem{petclinic}
\BIBentryALTinterwordspacing
``{Spring PetClinic Sample Application},'' 2024. [Online]. Available: \url{https://github.com/spring-projects/spring-petclinic}
\BIBentrySTDinterwordspacing

\bibitem{codamosa:artifact}
\BIBentryALTinterwordspacing
``{CodaMOSA Artifact},'' 2024. [Online]. Available: \url{https://github.com/microsoft/codamosa/tree/main/replication}
\BIBentrySTDinterwordspacing

\bibitem{jacoco}
EclEmma, ``{JaCoCo: Java Code Coverage Library},'' accessed: 2024-07-27.

\bibitem{tsdetect}
\BIBentryALTinterwordspacing
``{TSDetect},'' 2024. [Online]. Available: \url{https://github.com/TestSmells/TSDetect}
\BIBentrySTDinterwordspacing

\bibitem{peruma:2020:tsdetect}
A.~Peruma, K.~Almalki, C.~D. Newman, M.~W. Mkaouer, A.~Ouni, and F.~Palomba, ``tsdetect: an open source test smells detection tool,'' in \emph{Proceedings of the 28th ACM Joint Meeting on European Software Engineering Conference and Symposium on the Foundations of Software Engineering}, 2020, p. 1650–1654.

\bibitem{treesitter}
\BIBentryALTinterwordspacing
``{Tree-sitter},'' 2024. [Online]. Available: \url{https://tree-sitter.github.io/tree-sitter}
\BIBentrySTDinterwordspacing

\bibitem{wala}
\BIBentryALTinterwordspacing
``{WALA},'' 2024. [Online]. Available: \url{https://github.com/wala/WALA}
\BIBentrySTDinterwordspacing

\bibitem{tufano2020unit}
M.~Tufano, D.~Drain, A.~Svyatkovskiy, S.~K. Deng, and N.~Sundaresan, ``Unit test case generation with transformers and focal context,'' \emph{arXiv preprint arXiv:2009.05617}, 2020.

\bibitem{openaiapi}
\BIBentryALTinterwordspacing
OpenAI, ``Openai api,'' 2024. [Online]. Available: \url{https://platform.openai.com/docs/api-reference/introduction}
\BIBentrySTDinterwordspacing

\bibitem{ugare2024improving}
S.~Ugare, T.~Suresh, H.~Kang, S.~Misailovic, and G.~Singh, ``Improving llm code generation with grammar augmentation,'' \emph{arXiv preprint arXiv:2403.01632}, 2024.

\bibitem{ansible}
\BIBentryALTinterwordspacing
``{Ansible},'' 2024. [Online]. Available: \url{https://github.com/ansible/ansible}
\BIBentrySTDinterwordspacing

\bibitem{tornadoweb}
\BIBentryALTinterwordspacing
``{Tornado Web Server},'' 2024. [Online]. Available: \url{https://github.com/tornadoweb/tornado}
\BIBentrySTDinterwordspacing

\bibitem{flutes}
\BIBentryALTinterwordspacing
``{Flutes},'' 2024. [Online]. Available: \url{https://github.com/huzecong/flutes}
\BIBentrySTDinterwordspacing

\bibitem{codeql}
CodeQL, ``{CodeQL},'' accessed: 2024-07-27.

\bibitem{arcuri2019restful}
A.~Arcuri, ``Restful api automated test case generation with evomaster,'' \emph{ACM Transactions on Software Engineering and Methodology (TOSEM)}, vol.~28, no.~1, pp. 1--37, 2019.

\bibitem{pacheco2007randoop}
C.~Pacheco and M.~D. Ernst, ``Randoop: feedback-directed random testing for java,'' in \emph{Companion to the 22nd ACM SIGPLAN conference on Object-oriented programming systems and applications companion}, 2007, pp. 815--816.

\bibitem{tzoref2022tackletest}
R.~Tzoref-Brill, S.~Sinha, A.~A. Nassar, V.~Goldin, and H.~Kermany, ``Tackletest: A tool for amplifying test generation via type-based combinatorial coverage,'' in \emph{2022 IEEE Conference on Software Testing, Verification and Validation (ICST)}.\hskip 1em plus 0.5em minus 0.4em\relax IEEE, 2022, pp. 444--455.

\bibitem{wang2024software}
J.~Wang, Y.~Huang, C.~Chen, Z.~Liu, S.~Wang, and Q.~Wang, ``Software testing with large language models: Survey, landscape, and vision,'' \emph{IEEE Transactions on Software Engineering}, 2024.

\bibitem{vikram2023can}
V.~Vikram, C.~Lemieux, and R.~Padhye, ``Can large language models write good property-based tests?'' \emph{arXiv preprint arXiv:2307.04346}, 2023.

\bibitem{bareiss2022code}
P.~Barei{\ss}, B.~Souza, M.~d'Amorim, and M.~Pradel, ``Code generation tools (almost) for free? a study of few-shot, pre-trained language models on code,'' \emph{arXiv preprint arXiv:2206.01335}, 2022.

\bibitem{ryan2024code}
G.~Ryan, S.~Jain, M.~Shang, S.~Wang, X.~Ma, M.~K. Ramanathan, and B.~Ray, ``Code-aware prompting: A study of coverage guided test generation in regression setting using llm,'' \emph{FSE}, 2024.

\bibitem{schafer2023empirical}
M.~Sch{\"a}fer, S.~Nadi, A.~Eghbali, and F.~Tip, ``An empirical evaluation of using large language models for automated unit test generation,'' \emph{IEEE Transactions on Software Engineering}, 2023.

\bibitem{tufano2022generating}
M.~Tufano, D.~Drain, A.~Svyatkovskiy, and N.~Sundaresan, ``Generating accurate assert statements for unit test cases using pretrained transformers,'' in \emph{Proceedings of the 3rd ACM/IEEE International Conference on Automation of Software Test}, 2022, pp. 54--64.

\bibitem{yuan2024evaluating}
Z.~Yuan, M.~Liu, S.~Ding, K.~Wang, Y.~Chen, X.~Peng, and Y.~Lou, ``Evaluating and improving chatgpt for unit test generation,'' \emph{Proceedings of the ACM on Software Engineering}, vol.~1, no. FSE, pp. 1703--1726, 2024.

\bibitem{chen2024chatunitest}
Y.~Chen, Z.~Hu, C.~Zhi, J.~Han, S.~Deng, and J.~Yin, ``Chatunitest: A framework for llm-based test generation,'' in \emph{Companion Proceedings of the 32nd ACM International Conference on the Foundations of Software Engineering}, 2024, pp. 572--576.

\bibitem{pizzorno2024coverup}
J.~A. Pizzorno and E.~D. Berger, ``Coverup: Coverage-guided llm-based test generation,'' \emph{arXiv preprint arXiv:2403.16218}, 2024.

\bibitem{meta2023testingchallenges}
N.~Alshahwan, M.~Harman, and A.~Marginean, ``Software testing research challenges: An industrial perspective,'' in \emph{2023 IEEE Conference on Software Testing, Verification and Validation (ICST)}.\hskip 1em plus 0.5em minus 0.4em\relax IEEE, 2023, pp. 1--10.

\bibitem{meta2024observationbased}
N.~Alshahwan, M.~Harman, A.~Marginean, R.~Tal, and E.~Wang, ``Observation-based unit test generation at meta,'' in \emph{Companion Proceedings of the 32nd ACM International Conference on the Foundations of Software Engineering}, 2024, pp. 173--184.

\end{thebibliography}


\end{document}